\documentclass[12pt]{iopart}

\usepackage{graphicx}
\usepackage{xcolor}

\begin{document}

\title{Coupling of ``cold" electron plasma wave via stationary ion inhomogeneity to the plasma bulk} 

\author{Sanjeev Kumar Pandey, Jagannath Mahapatra and Rajaraman Ganesh}

\address{Institute for Plasma Research (IPR), Bhat, Gandhinagar 382428, India}
\address{Homi Bhabha National Institute (HBNI), Training School Complex, Anushaktinagar, Mumbai, Maharashtra 400094, India}
\ead{sanjeev.pandey@ipr.res.in}
\vspace{10pt}
\begin{indented}
\item[]May 2022
\end{indented}

\begin{abstract}
Using high resolution kinetic (VPPM-OMP 1.0) and fluid (BOUT++) solvers, evolution of long-wavelength electron plasma wave (EPW) in the presence of stationary periodic ion background non-uniformity is investigated. Mode coupling dynamics between long-wavelength EPW mode of scale $k$ and ion inhomogeneity of scale $k_{0}$ is illustrated. Validity of well known Bessel function $J_{n}(x)$ scaling in the cold plasma approximation (i.e., when phase velocity $\omega/k \gg v_{thermal}$) alongwith the effect of ion inhomogeneity amplitude ($A$) on temporal evolution of energy density in the long-wavelength EPW mode is investigated. Effect of finite system sizes on the Bessel $J_{n}(x)$ scaling is examined and scaling law for $\tau_{FM}$ i.e the time required to attain first minimum of energy density of the corresponding perturbed mode (also called phase mixing time for $k \longrightarrow 0$ modes) versus ion inhomogeneity amplitude $A$ obtained from both kinetic and fluid solutions for each of the cases studied, alongwith some major differences in $\tau_{FM}$ scaling for small system sizes is also reported.
\end{abstract}
%
\vspace{2pc}

\section{Introduction}\label{sec:Introduction}

      The excitation and evolution of electron plasma waves (EPW) has been a topic of extensive research for more than a century since their discovery by Tonks and Langmuir \cite{tonks_langmuir1929}. Pioneering work, in both linear and nonlinear plasma regimes have brought out several interesting physical aspects of electron plasma waves \cite{bohm1949,landau1946,kampen1955,bgk1957,dawson1959,oneil1965,kds1969}. One of the significant interest in this context is to study the evolution of long wavelength or high phase velocity EPWs in the presence of finite amplitude ion density modulation in the background \cite{kruer_kaw_1970,kruer_prl1970}. In their work, Kruer \cite{kruer1972} argued that the energy transfer from high phase velocity EPW modes into much slower ones and then into the particles via efficient off-resonant mode coupling is possible with finite amplitude ion inhomogeneity background. In realistic scenarios such as basic plasma laboratories, tokamaks, and astrophysical plasmas, equilibria are inhomogeneous in nature and hence such studies are also important to understand the involved underlying physics.

    In the past, attempts were made to understand the evolution and energy transfer mechanism of EPW modes in a bounded \cite{jackson1966,dorman1970} and periodic inhomogeneous equilibria \cite{bertrand_feix_baumann_1971}. In another interesting work, using warm plasma fluid model in one dimension (1D) with periodic boundaries, Kaw et al. \cite{kaw1973} illustrated energy transfer mechanism which led to ``fluid'' damping of high phase velocity EPW mode via mode coupling phenomenon with inhomogeneous stationary background of ions. They showed that a long-wavelength electron plasma wave of mode number $k$ can interact with an non-uniform, immobile ion background of scale $k_{0}$ to produce secondary modes with wavenumber $|k \pm Nk_{0}|$ (where $N$ is a dimensionless coupling parameter to be introduced later), resulting into cascading energy transfer upto $N^{th}$ sideband mode. In the process, the primary mode may lose its amplitude either fully or partially, depending on the value of $N$. Also, for cold plasma limit,  Kaw et al. \cite{kaw1973} argued that for very large system size $L \longrightarrow \infty$, the energy density i.e $|\delta E_{k}|^{2}$ of the amplitude of the modes with wavenumber $|k \pm Nk_{0}|:(N \longrightarrow 0,1,2,3...)$ will evolve in time $t$ according to the $N^{th}$ order Bessel function as $J^{2}_{N}(At/2)$ where $A$ is the amplitude of equilibrium ion inhomogeneity. These studies are proven useful in understanding various plasma concepts such as wave breaking in inhomogeneous equilibria \cite{sarkar2013,karmakar2018,xu2019}, laser plasma interaction \cite{everett1995,everett1996}, plasma instabilities \cite{Buchelnikova_1980,Buchelnikova_1981,barr1986,villeneuve1987,shukla_2009,Pandey_2021_TPI_2}, laser absorption by ion acoustic turbulence \cite{estabrook1981}, ionospheric turbulence \cite{pottelette1984,guzdar1996} etc. In present paper we shall ask ourselves as to how general is the Bessel function scaling for realistic, finite size systems with stationary ion inhomogeneity.
    
    Recently, in one of their works on linear Landau damping in a 1D periodic inhomogeneous plasmas, using Vlasov-Poisson model, Pandey et al. \cite{sanjeev2021} showed that time evolution of the energy density of the primary mode i.e $|\delta E_{k}|^{2}$ with $v_{\phi} \gg v_{the}$ (where $v_{\phi}=\omega/k$ is phase velocity of the perturbation, $v_{the}$ is the electron thermal velocity), does not evolve according to Bessel approximation $J_{0}^{2}(x)$ \cite{kaw1973}, for system size $L = 20 \pi \lambda_{D}$ such that $v_{\phi} \sim 10 v_{the}$ (still within fluid approximation). In the present work, in order to ascertain the generality of the Bessel function scaling, extensive study for various system sizes $L$, inhomogeneity scales $A$ at infinitesimal perturbation (or linear) amplitude $\alpha$ becomes necessary. Also, the interesting question of the role of equilibrium electric field due to inhomogeneity, though small in magnitude, is also addressed. In light of the above studies \cite{kaw1973,sanjeev2021}, we have used both the warm fluid simulation approach which will provide an exact solution for warm fluid equation with an equilibrium background electric field as well as kinetic simulation approach for a exact one on one comparative study. 
 
  In this paper, using high resolution OpenMP based kinetic (VPPM-OMP 1.0) Vlasov-Poisson as well as fluid (BOUT++) solvers, for identical initial conditions, we numerically solve kinetic and fluid model equations respectively to address the dynamics of long-wavelength EPW modes in the presence of non-uniform periodic ion background inhomogeneity for various system sizes and inhomogeneity amplitudes and inhomogeneity scale lengths. We first demonstrate for large systems sizes that the damping and energy transfer mechanism of long-wavelength EPW modes due to mode coupling phenomenon with background ion density is exactly analogous to Kaw et al. \cite{kaw1973}. Quantitative comparative study between kinetic and fluid solutions are performed to investigate the validity of Bessel scaling approximation and to understand the impact of kinetic effects on the solutions. Also, we bring out the effect of ion inhomogeneity amplitude $A$ on the evolution of these EPW modes as well as in the estimation of time for the first minimum of energy density of primary modes to occur (say $\tau_{FM}$). Finally, investigations on the Bessel scaling of energy density in EPW modes as a function of time, with finite system sizes (L) are also performed and $\tau_{FM}$ scaling using both warm fluid and kinetic approaches, is reported.

   The rest of this paper is organized as follows: In Secs. \ref{Fluid Model} and \ref{Kinetic Model}, warm fluid and kinetic model equations is presented. We discuss numerical scheme for both fluid and kinetic solvers in Sec. \ref{Numerical Scheme}. We present the simulation results obtained for inhomogeneous plasma case with $k_{min}= 0.00625$ (in Sec. \ref{Simulation Results Inhomogeneous plasma case with kmin}), finite inhomogeneity amplitude effects (in Sec. \ref{Finite inhomogeneity amplitude effects}) and finite size effects (in Sec. \ref{Finite Size effects}). Finally, we conclude in Sec. \ref{Discussion and conclusion}.
   
\section{Mathematical Model and Numerical Scheme }
\subsection{ Kinetic Model }
\label{Kinetic Model}

  In the kinetic theory framework, small amplitude electron plasma wave (EPW) is modeled in an unmagnetized, collisionless, spatially inhomogeneous plasma system consisting of ions and electrons by a set of coupled one dimensional Vlasov-Poisson (VP) equations [\cite{raghunathan2013,pallavi2016,pallavi2017,pallavi2018,pallavi2020,Saini2018,pallavithesis,sanjeev2021,Pandey_2021_TPI_1,Pandey_2021_TPI_2}],    

\begin{equation}
\frac{\partial f_{e}}{\partial t}+v_{e}\frac{\partial f_{e}}{\partial x}-E\frac{\partial f_{e}}{\partial v_{e}}=0
\label{EQ_ELE_VLASOV}
\end{equation}

\begin{equation}
\frac{\partial f_{i}}{\partial t}+v_{i}\frac{\partial f_{i}}{\partial x}+\frac{E}{m_{r}}\frac{\partial f_{i}}{\partial v_{i}}=0
\label{EQ_ION_VLASOV}
\end{equation}

\begin{equation}
\frac{\partial E}{\partial x}= \int f_{i}dv_{i} - \int f_{e}dv_{e}
\label{EQ_POISSON}
\end{equation}
where $f_{i}(x,v,t) $ and $f_{e}(x,v,t) $ is the ion and electron distribution function respectively, $m_{r}$ is the mass ratio of ions to electrons, i.e, $m_{r} = M_{i} /M_{e}$, $E=E(x,t)$ is the total electric field. Eq. \ref{EQ_ELE_VLASOV} - \ref{EQ_POISSON} time $t$ is normalized to electron plasma frequency $\omega_{pe}^{-1}$, spatial coordinate $x$ is normalized to electron Debye length $\lambda_{D}$, velocities to electron thermal velocity $v_{the}=\lambda_{D}\omega_{pe}^{-1}$, electric field to $en_{0}\lambda_{D}/ \epsilon_{0}$, and distribution function has been normalized to $n_{0}/\lambda_{D}\omega_{pe}$ where $n_{0}$ is uniform plasma density.

 Considering Maxwellian velocity distribution function $f_{Me}(v_{e})$ for electrons with an arbitrary spatial inhomogeneity $n_{0e}(x)$, the equilibrium solution of Vlasov equation requires (prime denotes spatial differentiation),

\begin{equation}
E_0(x)= - \left[ \frac{n_{0e}^{'}(x)}{n_{0e}(x)} \right]
\label{EQ_EOX}
\end{equation}
where $E_{0}(x)$ is the equilibrium electric field resulting from equilibrium inhomogeneity \cite{dorman1970}. In this model, ions form a nonuniform, stationary $(\partial/\partial t=0)$ background of number density $n_{0i}(x)$ which is determined by satisfying the equilibrium Poisson's equation,

\begin{equation}
\frac{\partial E_{0}(x)}{\partial x}= \int f_{0i}dv_{i} - \int f_{0e}dv_{e}
\label{EQ_EPE}
\end{equation}
Combining Eqs. \ref{EQ_EOX}, \ref{EQ_EPE} and using equilibrium Maxwellian velocity distribution for electrons and ions alongwith their respective spatial inhomogeneities i.e $n_{0e}(x)$ and $n_{0i}(x)$ one obtains, 

\begin{equation}
\frac{\partial}{\partial x} \left [ - \frac{n_{0e}'(x)}{n_{0e}(x)} \right] = \int n_{0i}(x)f_{Mi}(v_{i}) dv_{i} - \int n_{0e}(x)f_{Me}(v_{e}) dv_{e}
\label{EQ_EPE_2}
\end{equation}
which results in,

\begin{equation}
n_{0i}(x)= n_{0e}(x) + \left[ \frac{(n_{0e}^{'}(x))^{2}-n_{0e}^{''}(x)n_{0e}(x)}{n_{0e}^{2}(x)} \right]
\label{EQ_NOI_1}
\end{equation}
Thus, using Eq. \ref{EQ_ELE_VLASOV} - \ref{EQ_POISSON}, spatial inhomogeneous equilibrium density profile for ions $n_{0i}(x)$ and for electrons $n_{oe}(x)$ are obtained self-consistently, taking into account the equilibrium electric field variation given by Eq. \ref{EQ_EOX} . Consequently, one can presume any periodic density profile for $n_{0e}(x)$ and obtain the corresponding $E_{0}(x)$ and $n_{0i}(x)$ from Eq. \ref{EQ_EOX} and \ref{EQ_NOI_1} respectively (or vice versa). Note that if $n_{0e}^{''}(x)=n_{0e}^{'}(x)=0$, from Eq. \ref{EQ_EOX} and \ref{EQ_NOI_1} we obtain $E_{0}(x)=0$ and $n_{0i}(x)=n_{0e}(x)$ respectively \cite{sanjeev2021}.
 
\begin{figure*}
\centerline{\includegraphics[scale=0.37]{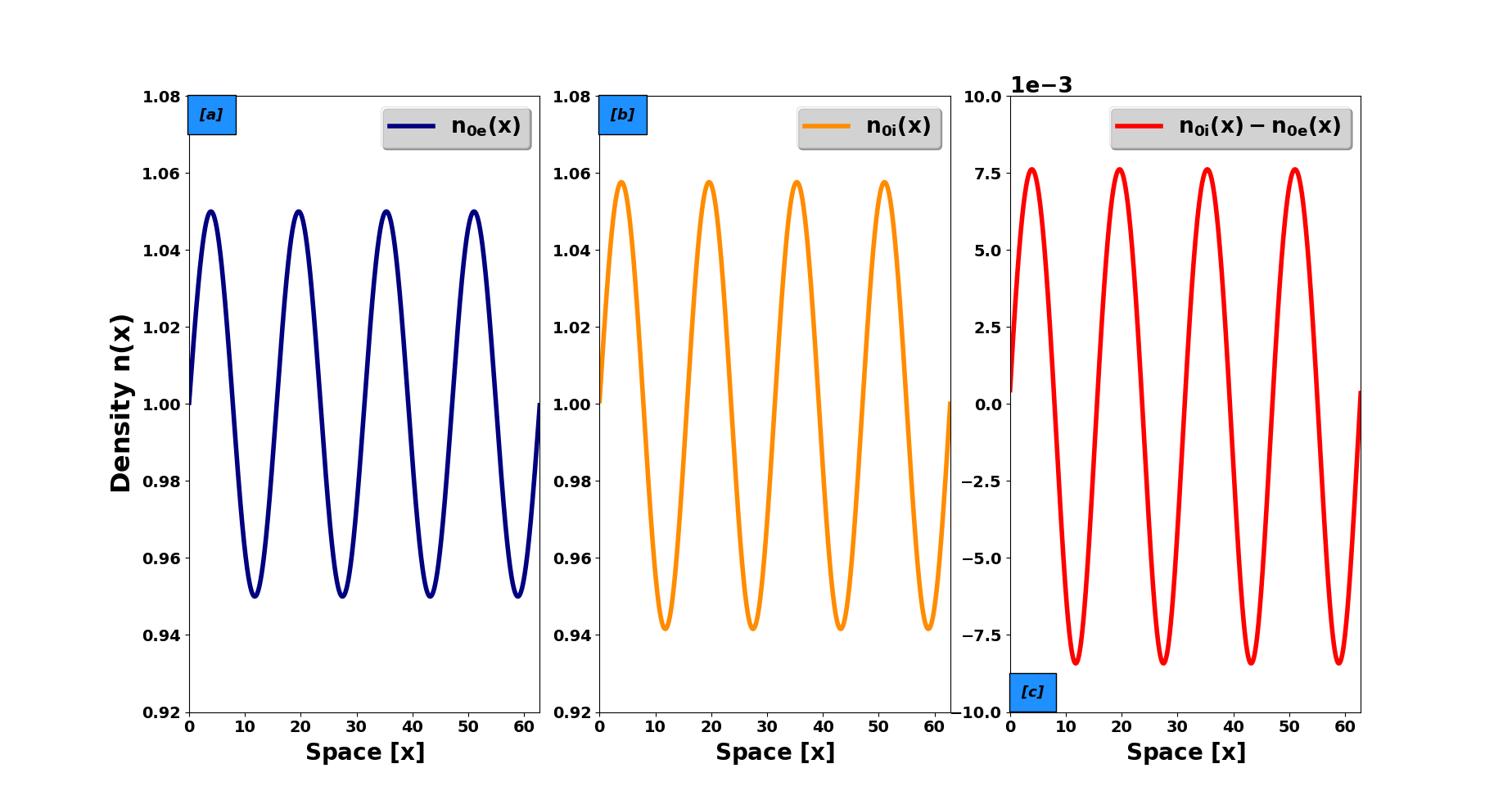}}
\caption{ Variation of spatially inhomogeneous equilibrium density profiles i.e (a) $n_{0e}(x)$, (b) $n_{0i}(x)$ (defined by Eq. \ref{EQ_NOE} and \ref{EQ_NOI}) and (c) $n_{0i}(x)-n_{0e}(x)$ with respect to space $(x)$ for $A=0.05,~k_{0}=4k_{min},~k_{min}=0.1$ case with $L_{max}=2\pi/k_{min}=62.8318$.}
\label{41_NOE+NOI_A=0.05_KMIN=0.1}
\end{figure*}

  Let us consider a simple periodic form for the spatially inhomogeneous density profiles $n_{0e}(x)$ as,
  
\begin{equation}
n_{0e}(x)= 1+A \sin (k_{0}x)
\label{EQ_NOE}
\end{equation}
Using Eq. \ref{EQ_EOX} , \ref{EQ_NOI_1} and \ref{EQ_NOE} one obtains $n_{0i}(x)$, $E_{0}(x)$ as,

\begin{equation}
n_{0i}(x)= \left[ \frac{(1+A \sin (k_{0}x))^{3}+A^{2}k_{0}^{2}+Ak_{0}^{2}\sin (k_{0}x) }{1+A^{2}\sin^{2}(k_{0}x)+2A\sin(k_{0}x)} \right]
\label{EQ_NOI}
\end{equation} 

\begin{equation}
E_{0}(x)=- \left[ \frac{(Ak_{0}\cos(k_{0}x))}{(1+A\sin(k_{0}x))} \right]
\label{EQ_E0}
\end{equation}
where $k_{0}$ is the measure of equilibrium inhomogeneity scale expressed in integer multiples of $k_{min}$ where $k_{min}=2\pi/L_{max}$ and $A$ is the strength of equilibrium inhomogeneity. As may be expected, when the value of  $A$ is set to zero, $E_0(x)=0$ and $n_{0i}(x)$ becomes homogeneous in space. It is important to note that periodicity is enforced in the system, which is essential and necessary condition for consistently addressing periodic non-uniform profiles described by Eq. \ref{EQ_NOE}, \ref{EQ_NOI}. Hence, every equilibrium inhomogeneity scale $k_{0}$ and all the perturbation $k$ scales in the simulation are expressed as integer multiples of $k_{min}$. In Fig. \ref{41_NOE+NOI_A=0.05_KMIN=0.1} [(a) and (b)], Variation of spatially inhomogeneous equilibrium density profiles i.e $n_{0e}(x)$, $n_{0i}(x)$ (defined by Eq. \ref{EQ_NOE} and \ref{EQ_NOI} respectively) and $n_{0i}(x)-n_{0e}(x)$ with respect to space $(x)$ are shown for $A=0.05,~k_{0}=4k_{min},~k_{min}=0.1$ case with $L_{max}=2\pi/k_{min}=62.8318$.

\subsection{ Warm Fluid Model }
\label{Fluid Model}

    Using 1D warm fluid theory, a small amplitude electron plasma wave (EPW) in a plasma with finite ion density modulation in the background can be studied by considering a stationary self-consistently obtained ion density equilibrium profile from Eq. \ref{EQ_NOI},

\begin{equation}
n_{i}=n_{0} \left[ \frac{(1+A \sin (k_{0}x))^{3}+A^{2}k_{0}^{2}+Ak_{0}^{2}\sin (k_{0}x) }{1+A^{2}\sin^{2}(k_{0}x)+2A\sin(k_{0}x)} \right]
\label{EQ_NI_FLUID}
\end{equation}
where $A$ and $k_{0}$ are the amplitude and scale of the background ion density profile respectively. 
  
   The corresponding one dimensional warm fluid equations i.e continuity and equation of motion for electrons are \cite{kaw1973}, 

\begin{equation}
\frac{\partial n}{\partial t} + \frac{\partial }{\partial x}(nv)=0
\label{EQ_CONTINUITY_FLUID}
\end{equation}

\begin{equation}
\frac{\partial}{\partial t}(nv) + \frac{\partial }{\partial x}(nv^{2})=-\frac{neE}{m}-\frac{\gamma KT}{m} \left[ \frac{\partial n}{\partial x} \right]
\label{EQ_EOM_FLUID}
\end{equation}
and the Poisson's equation,

\begin{equation}
\frac{\partial \delta E}{\partial x}=-4 \pi e \delta n
\label{EQ_POISSON_FLUID}
\end{equation}
where

\begin{equation}
n= n_{0e}(x) + \delta n
\label{EQ_LIN_N_FLUID}
\end{equation}

\begin{equation}
v= v_{0} + \delta v ~:~ v_{0}=0
\label{EQ_LIN_VEL_FLUID}
\end{equation}

\begin{equation}
E= E_{0} + \delta E ~:~ E_{0} \neq 0
\label{EQ_LIN_EF_FLUID}
\end{equation}
where $n_{0e}(x)=n_{0}(1+A \sin (k_{0}x))$, $K$ is the Boltzmann's constant, $\gamma$ is the ratio of specific heats, $\delta n$ is the perturbation of the electron density from the ion density and $\delta v$ is the perturbation in the electron velocity and equilibrium electron velocity is assumed to be zero i.e $v_{0}=0$. Here, $E_{0}$ is the non-zero finite equilibrium electric field due to presence of the background ion inhomogeneity given by Eq. \ref{EQ_E0} and $\delta E$ is the perturbation in the electric field. Linearizing Eq. \ref{EQ_CONTINUITY_FLUID} and \ref{EQ_EOM_FLUID} in the perturbations, one obtains,

\begin{equation}
\frac{\partial \delta n}{\partial t} + \frac{\partial}{\partial x} \left[ n_{0}(1+A \sin (k_{0}x)) \delta v \right] = 0
\label{EQ_FINAL_1_FLUID}
\end{equation}

\begin{equation}
\frac{\partial}{\partial t} \left[ n_{0}(1+A \sin (k_{0}x)) \delta v \right]= - \frac{e}{m} \left[ n_{0}(1+A \sin (k_{0}x)) \delta E + E_{0} \delta n \right] -\frac{\gamma KT}{m} \left[ \frac{\partial \delta n}{\partial x} \right]
\label{EQ_FINAL_2_FLUID}
\end{equation}
Combining Eqs. \ref{EQ_FINAL_1_FLUID} and \ref{EQ_FINAL_2_FLUID} results,  

\begin{equation}
\frac{\partial^{2} \delta n}{\partial t^{2}} - \left[ \frac{\omega_{p}^{2}}{4 \pi e} \right] \frac{\partial}{\partial x}[(1+A \sin (k_{0}x))\delta E]-\frac{e}{m} \frac{\partial}{\partial x} [E_{0} \delta n] -\frac{\gamma KT}{m} \left[ \frac{\partial^{2} \delta n}{\partial x^{2}} \right] = 0
\label{EQ_FINAL_3_FLUID}
\end{equation}
where $\omega_{p}=(4\pi n_{0}e^{2}/m)^{1/2}$ is the mean plasma frequency of the uniform plasma. Using Eq. \ref{EQ_POISSON_FLUID} and eliminating $\delta n$ from Eq. \ref{EQ_FINAL_3_FLUID} results in,

\begin{equation}
\frac{\partial^{2} \delta E}{\partial t^{2}} + \omega_{p}^{2}[1+A \sin (k_{0}x)]\delta E-\frac{e E_{0}}{m} \left[ \frac{\partial \delta E}{\partial x} \right] -\frac{\gamma KT}{m} \left[ \frac{\partial^{2} \delta E}{\partial x^{2}} \right] = 0
\label{EQ_FINAL_4_FLUID}
\end{equation}
Eq. \ref{EQ_FINAL_4_FLUID} illustrates the basic equation for plasma oscillations in the presence of a background stationary equilibrium ion density profile and an equilibrium electric field $E_{0}$. If time is normalized to $\omega_{p}^{-1}$, space is normalized to $\lambda_{D}$ corresponding to a homogeneous plasma density, one obtains,  

\begin{equation}
\frac{\partial^{2} \delta E}{\partial t^{2}} + [1+A \sin (k_{0}x)]\delta E+ E_{0} \left[ \frac{\partial \delta E}{\partial x} \right] - \gamma \left[ \frac{\partial^{2} \delta E}{\partial x^{2}} \right] = 0
\label{EQ_FINAL_5_FLUID}
\end{equation}
Eq. \ref{EQ_FINAL_5_FLUID} represents the normalized equation for plasma oscillations in the presence of non-uniform, immobile background of ions with a finite non-zero background equilibrium electric field $E_{0}(x)$ given by Eq. \ref{EQ_E0} of the kinetic model as,

\begin{equation}
E_{0}(x)=- \left[ \frac{(Ak_{0}\cos(k_{0}x))}{(1+A\sin(k_{0}x))} \right]
\label{EQ_FINAL_6_FLUID}
\end{equation}
Eq. \ref{EQ_FINAL_5_FLUID} reduces to Eq. (5) of Kaw et al. \cite{kaw1973} for $E_{0}=0$, which was then solved analytically in Ref. \cite{kaw1973} by rewriting the same into a Mathieu equation. The modulation of energy of a long wavelength EPW (i.e, $k \longrightarrow 0$ or $L \longrightarrow \infty$, such that $\omega/k \gg v_{thermal}$) via fluid model and its coupling to background inhomogeneity $k_{0}$ was addressed analytically using Bessel function based solutions. In this work, we shall revisit these findings by solving Eq. \ref{EQ_FINAL_5_FLUID} numerically without making any approximation and compare the same with a fully kinetic solutions, for a range of interesting parameters.

\subsection{Numerical Schemes}
\label{Numerical Scheme}

    For numerical simulation of warm fluid model, we solve Eq. \ref{EQ_FINAL_5_FLUID} using BOUT++ framework, which is an open-source, nonlinear, finite difference based plasma simulation code written in C++ language. It has a highly modular structure and in-built MPI parallelization, enabling it to solve any number of coupled partial differential equations (PDEs) in both slab and curvilinear geometries \cite{dudson2009bout++, dudson2015bout++}. Recently, BOUT++ framework has been used for plasma simulation for a variety of theoretical models such as reduced-magnetohydrodynamics (RMHD) model \cite{mahapatra2021}, gyro-fluid model \cite{xu_2013}, kinetic-fluid closure models [such as Landau-fluid (LF) closure] \cite{libo_wang_2019, ben_zhu_2021}, Vlasov simulations \cite{tavassoli2021role} etc. Here we solve the Eq. \ref{EQ_FINAL_5_FLUID} in 1D by splitting it into two coupled first order differential equations in time, with periodic boundaries. The CVODE time solver (implicit type) is used to evolve the equations in time with a time step 0.1 $\omega_{pe}^{-1}$. The spatial domain size $D=[0,L_{max}]$ and grid points $N_{x}$ for each simulation set are taken exactly same as Vlasov simulation sets as tabulated in Table \ref{TABLE 1}.     

\begin{table}
\caption{ Parameter set i.e $k_{min},~L_{max},~N_{x},~\Delta x,~T_{R}$ for the fluid (BOUT++) and kinetic (VPPM - OMP 1.0) solvers. Here, $\Delta v=2v_{max}/N_{v}$ and $k=k_{min}$. }  
\centering                         
\begin{tabular}{c c c c c c}           
\hline\hline                        
$k_{min}~~$ & $L_{max}~~~$ & Bout++ $~~$ & VPPM - OMP 1.0 $~$ & $\Delta x$ & $T_{R}$  \\ 
$~$ & $(2\pi/k_{min})$ & $(N_{x})$ & $(N_{x},~N_{v})$ & $(L_{max}/N_{x})$ & $(2\pi/(k \Delta v))$  \\[1.0ex]   
\hline            
0.1 & $20\pi$ & 2048 & 2048, 10000 & 0.030679 & 52359.8776  \\          
0.025 & $80\pi$ & 8192 & 8192, 10000 & 0.030679 & 209439.5102 \\
0.0125 & $160\pi$ & 16384 & 16384, 10000 & 0.030679 & 418879.0205 \\
0.00625 & $320\pi$ & 32768 & 32768, 10000 & 0.030679 & 837758.0410 \\ [1ex]
\hline                               
\end{tabular}
\label{TABLE 1}
\end{table}

    We have solved kinetic Eqs. \ref{EQ_ELE_VLASOV} - \ref{EQ_POISSON} numerically using VPPM-OMP 1.0 \cite{sanjeev2021,Pandey_2021_TPI_1,Pandey_2021_TPI_2}. It is an OpenMP based 1D Eulerian Vlasov-Poisson solver capable of handling both electron and ion dynamics. It is based on Piecewise Parabolic Method (PPM) advection scheme proposed by Colella and Woodward \cite{colella1984} and uses time stepping method given by Cheng and Knorr \cite{cheng1976}. A Fourier transform (FT) based method has been implemented for solving Poisson equation. Simulation domain in 1D phase space $(x,v)$ is set as $D=[0,L_{max}] \times [-v_{e}^{max},v_{e}^{max}]$, where $L_{max}=2\pi/k_{min}$ is the system size and $v_{e}^{max}=6.0$ chosen sufficiently large so that electron distribution function approaches zero as $|v|$ approaches $v_{e}^{max}$. The simulation domain is discretized into $N_{x}$ grid points in spatial domain and $N_{v}$ grid points in velocity domain for both ions and electrons [See Table \ref{TABLE 1} for details]. Periodic boundary conditions (PBC) have been implemented in both spatial and velocity domains.
 
     Simulation is initialized, with normalized Maxwellian distribution function for ions and electrons with respective spatial inhomogeneous equilibrium density profiles $n_{0i}(x)$, $n_{0e}(x)$ obtained earlier and linear cosinusoidal density perturbation given as,

\begin{equation}
f_{e}(x,v_{e},t=0)= (n_{0e}(x)+\alpha cos(kx))f_{Me}(v_{e})
\label{EQ_FE_T=0}
\end{equation}

\begin{equation}
f_{i}(x,v_{i},t=0)= n_{0i}(x)f_{Mi}(v_{i})
\label{EQ_FI_T=0}
\end{equation}

\begin{equation}
f_{Me}(v_{e})= \frac{1}{\sqrt{2\pi}} exp \left[ \frac{-v_{e}^{2}}{2} \right]
\label{EQ_FE_MAX}
\end{equation}

\begin{equation}
f_{Mi}(v_{i})= \frac{1}{\sqrt{2\pi}}\sqrt{\frac{m_{r}}{T_{r}}} exp \left[ \frac{-m_{r}v_{i}^{2}}{2T_{r}}\right] 
\label{EQ_FI_MAX}
\end{equation}
where $f_{Mi}(v_{i})$, $f_{Me}(v_{e})$ are the normalized ion and electron Maxwellian velocity distribution functions, $T_{r}=T_{i}/T_{e}$ is temperature ratio of ions to electrons, $ m_{r}=M_{i}/M_{e}$ is mass ratio of ions to electrons, $\alpha $ is the strength or amplitude of perturbation, $k$ is the perturbation scale, always used as integer multiple of $k_{min}$. In the following, we present fluid and kinetic simulation results of long wavelength electrons plasma waves (EPWs) in the presence of background inhomogeneity with $m_{r} \gg 1$, considered for stationary ions. 

\section{Simulation Results Using Fluid And Kinetic Approaches}
\label{Simulation Results Using Fluid And Kinetic Approach}

  In this section, we will present the fluid and kinetic simulation results for inhomogeneous plasma cases with $k_{min}=0.00625$ alongwith the finite system size and finite ion inhomogeneity amplitude effects on the dynamics of the long-wavelength EPW modes.

\subsection{ Inhomogeneous plasma case with $k_{min}=0.00625$}
\label{Simulation Results Inhomogeneous plasma case with kmin}

\begin{figure*}
\centerline{\includegraphics[scale=0.36]{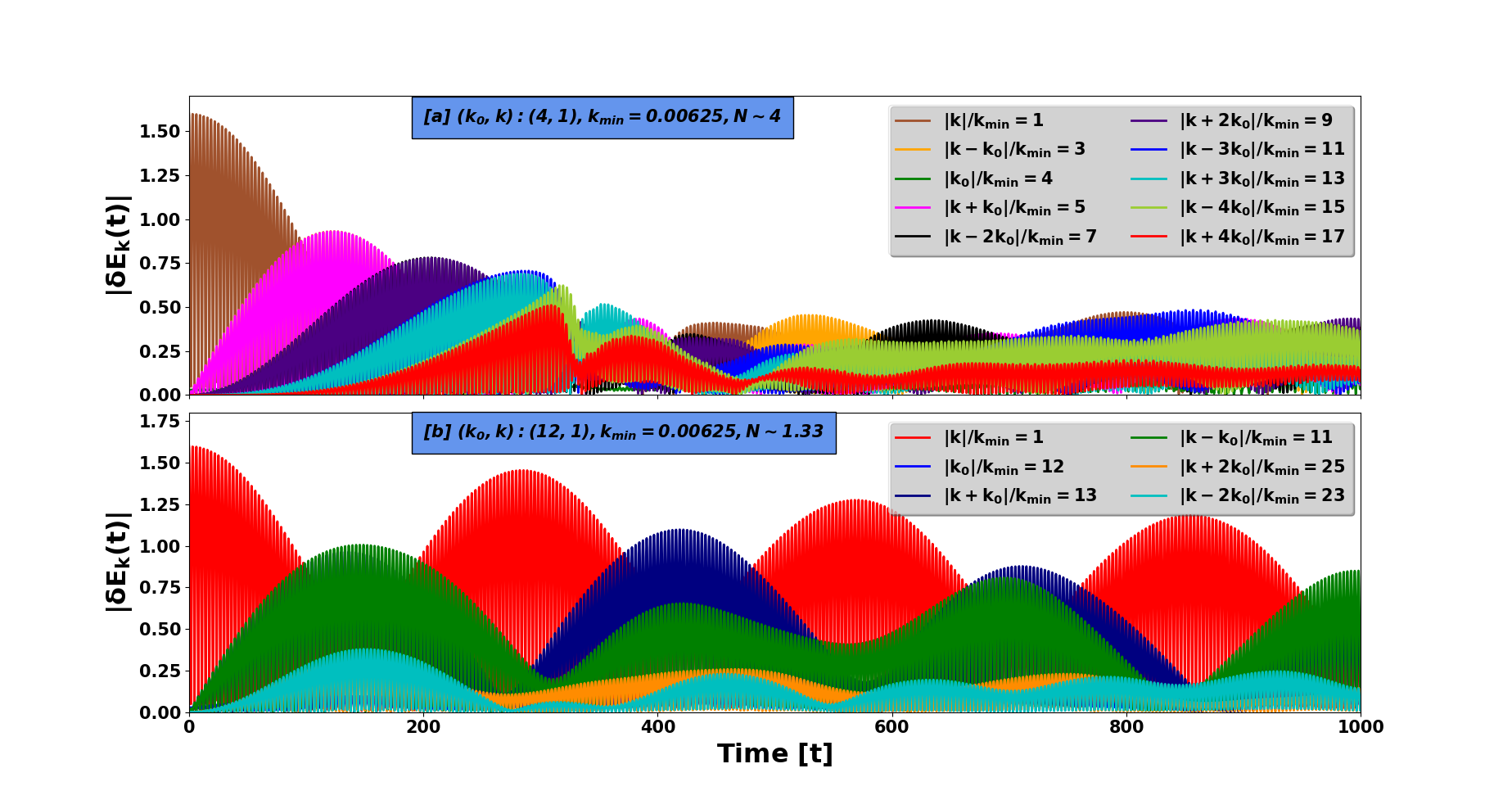}}
\caption{ Time evolution of perturbed $|k|$ and coupled $|k \pm Nk_{0}|$ electric field $|\delta E_{k}|$ modes obtained from kinetic solver with inhomogeneity amplitude $A=0.03$ and linear perturbation amplitude $\alpha=0.01,~k_{min}=0.00625,~N_{x} \times N_{v} = [32768 \times 10000]$ for (a) $(k_{0},k):(4,1)$ and (b) $(k_{0},k):(12,1)$. In (a), for $k_{0}=0.025,\gamma=3.0$, coupling parameter $N \simeq 4$ (estimated using $N^2 \sim A/(\gamma k_0^2)$ of Ref. \cite{kaw1973}), whereas for $k_{0}=0.075,\gamma=3.0$ in (b) coupling parameter $N$ is 1.33. Portrait (a) and (b) demonstrates the reduction in coupling of the sideband modes with increase in the inhomogeneity scale $k_{0}$ with fixed $A$ and $\alpha$. Also, it indicates that when the higher sidebands are not coupled efficiently, the primary mode recovers most of its initial energy in the subsequent cycles.}
\label{41_121_EK_ALPHA=0.01_A=0.03_kmin=0.00625}
\end{figure*}

      As previously mentioned, simulations are carried out in the limit of kinetic electrons and immobile inhomogeneous background of ions i.e $A \neq 0$ case. To begin with, in the kinetic (VPPM-OMP 1.0) solver, we initialize two spatial scales i.e equilibrium inhomogeneity and perturbation $(k_{0},k)$ scales respectively, from the beginning at $t=0~\omega_{pe}^{-1}$. Both the scales are expressed in integer multiples of $k_{min}$ which is set to 0.00625, in order to excite long-wavelength modes identical to the ones addressed in Ref. \cite{kaw1973} such that $L_{max}=2\pi/k_{min}=320\pi$. To facilitate both qualitative and quantitative comparison (shown later), we consider exactly identical parameter as Ref. \cite{kaw1973} i.e $(k_{0},k) = (m_{0}k_{min},n_{0}k_{min}) = (4,1)k_{min}$ with equilibrium inhomogeneity and linear perturbation amplitudes as $A=0.03$ and $\alpha=0.01$ respectively. Maximum electron velocity is set to $|v_{e}^{max}|=6.0$. Phase space grid resolutions are set to large values i.e $N_{x} \times N_{v}=32768 \times 10000$ to sufficiently resolve the $(x,v)$ domains. It is important to note that, recurrence time defined as $T_{R}=2\pi/(k \Delta v):k=k_{min}$ where $\Delta v=2v_{max}/N_{v}$ is tabulated in Table. \ref{TABLE 1} \cite{arber_vann2002,vannthesis,raghunathan2013,sanjeev2021}, which is much larger than any physically relevant time scale in the system. Similarly, in the 1D warm fluid (BOUT++) solver, we initiate the exact parameter set as mentioned above the details of which is discussed in Sec. \ref{Numerical Scheme}. In both kinetic and warm fluid solvers, simulations are advanced upto $t=1000~\omega_{pe}^{-1}$ in time. Evolution of Fourier mode from kinetic solver, after removing the equilibrium contributions of scale $k_{0}$ i.e $k : \delta E_{k}(t)$ is shown in Fig. \ref{41_121_EK_ALPHA=0.01_A=0.03_kmin=0.00625}  which is Fourier transform of $\delta E(x,t)$ defined as,

\begin{equation}
 \delta E(x,t)=E(x,t) - E_{0}(x)= \sum_{k} \delta E_{k}(t) e^{-ikx}
\label{EQ_DELTA_EF}
\end{equation}
where $E(x,t)$ is the total electric field obtained by solving Poisson’s equation Eq. \ref{EQ_POISSON}, $E_{0}(x)$ is the equilibrium electric field Eq. \ref{EQ_E0}, which for $A=0$ is $E_{0}=0$.

It is well known that in an homogeneous plasma, long-wavelength perturbations of an electron plasma wave (EPW) suffer no kinetic damping at all, since $v_{\phi}(=\omega/k)>>v_{the}$ i.e kinetic or Landau resonance conditions are not satisfied \cite{sanjeev2021}. In their study, using a warm plasma fluid model with 1D sheet simulations, Kaw et al. \cite{kaw1973} showed that ``fluid'' damping of such long-wavelength EPW mode is possible in the presence of an inhomogeneous background of ions. In both the studies \cite{kaw1973,sanjeev2021}, it was illustrated that a long-wavelength EPW mode of scale $k$ interacts with an inhomogeneous background of ions with scale $k_{0}$ to produce coupled modes of wavenumber $|k \pm Nk_{0}|$ where $N$ is the coupling parameter which may be estimated as \cite{kaw1973},    

\begin{equation}
N^{2} \sim \left[ \frac{A}{\gamma k_{0}^{2}} \right]
\label{EQ_N}
\end{equation}
and $\gamma$ is the ratio of specific heats, $k_{0}$ and $A$ are inhomogeneity scale and amplitude respectively. In the $k_{0}>>k$ limit, mode coupling of $k_{0}$ and $k$ scales results in coupling of high phase velocity mode with $v_{\phi}=\omega/k$ to low phase velocity oscillations at $v_{\phi}=\omega_{[k \pm Nk_{0}]}/(|k \pm Nk_{0}|)$ and high phase velocity mode (long-wavelength EPW) is made to interact with bulk of the plasma, leading to damping of energy in the primary mode. In a kinetic model, if the phase velocity of these secondary modes can resonantly interact with the plasma bulk, where the equilibrium electron velocity gradient $\partial f_{0}/ \partial v $ is finite, it may lead to kinetic effects such as wave particle energy exchange and hence damping of the secondary mode. It is important to note that the inhomogeneous background induces two levels of energy loses to the primary EPW mode. One is the energy transfered to coupled sideband $|k \pm Nk_{0}|$ modes which has no kinetic effects involved and secondly, energy transfer of secondary modes to thermal energy of particles, due to the Landau damping via resonant kinetic process if the phase velocity of the secondary modes are able to sample $\partial f_{0}/\partial v$ in the bulk plasma. This later process, has not been studied in detailed, before the advent of the present work.

  Fig. \ref{41_121_EK_ALPHA=0.01_A=0.03_kmin=0.00625} shows the time evolution of the perturbation as well as the coupled sideband mode electric field amplitudes $|\delta E_{k}|$ with respect to time for (a) $(k_{0},k):(4,1)$ and (b) $(k_{0},k):(12,1)$ pairs with $A=0.03,~\alpha=0.01,~k_{min}=0.00625$. For (a) $(k_{0},k):(4,1)$ pair, from Eq \ref{EQ_N}, coupling parameter $N \sim \sqrt{0.03/3(0.025)^{2}} \sim 4.0$, which indicates that the generated coupled sideband modes should be at $|k \pm Nk_{0}|=3,5,7,9,11,13,15,17:N \longrightarrow 1-4$, while for (b) $(k_{0},k):(12,1)$ pair, $N \sim \sqrt{0.03/3(0.075)^{2}} \sim 1.33$, $|k \pm Nk_{0}|=13,11,25,23$ sideband modes are expected to be generated. It is evident from Fig. \ref{41_121_EK_ALPHA=0.01_A=0.03_kmin=0.00625} [(a) and (b)] that in a non-dissipative system with three-mode coupling $(E_{k},~E_{k \pm k_{0}},~E_{k \pm 2k_{0}})$, energy will transfer back and forth between perturbed primary and coupled sideband modes and most of the initial energy will be recovered by the perturbed primary mode in the second cycle while in the $N^{th}$ mode coupling process i.e $N=4: (E_{k},~E_{k \pm k_{0}},~E_{k \pm 2k_{0}},~ E_{k \pm 3k_{0}},~ E_{k \pm 4k_{0}})$, the driven modes at $|k \pm (N-1)k_{0}|$ use part of its energy to excite the higher sideband modes at $|k \pm Nk_{0}|$. Therefore, in this case the perturbed mode $|E_{k}|$ will not be able to recover its initial energy in the subsequent cycles. In other words, if the higher sidebands are coupled, the primary perturbation mode $|E_{k}|$ will regain less of its energy in presence of background ion inhomogeneity.

   Kaw et al. \cite{kaw1973} illustrated the process of this mode coupling with a discussion on the solution of Mathieu equation for plasma oscillations in the presence of a background ion density inhomogeneity and proposed that as long as the frequency mismatch between perturbed and the $N^{th}$ sideband modes is less than the variation in plasma frequency associated with ion density fluctuation $[( \omega_{k} - \omega_{k \pm Nk_{0}}) \le \Delta \omega_{p0}]$, cascading energy transfer can be expected upto $N^{th}$ sideband mode. Also, for a cold plasma approximation (i.e, $v_{\phi}=\omega/k >> v_{the}$), these Authors argued that for very large system size i.e., for $L \rightarrow \infty$ or $k_{min} \rightarrow 0$, the amplitude of the mode with wavenumber $|k \pm Nk_{0}|:(N \longrightarrow 0,1,2,3...)$ will evolve in time according to the $N^{th}$ order Bessel function as $J_{N}(At/2)$. It is important to note that the earlier said Bessel function scaling should be expected to match in the limit of very large system size. One of the interesting questions, to be discussed here is, how would the primary mode amplitudes in time $t$ evolve, if the system size $L$ is sufficiently large but is finite, such that the phase velocity of the longest wavelength mode $v_{\phi} = \omega/k$ is much greater than $v_{the}$. We shall come back to this question later.

\begin{figure}
$~~~~~~~~$
\includegraphics[scale=0.27]{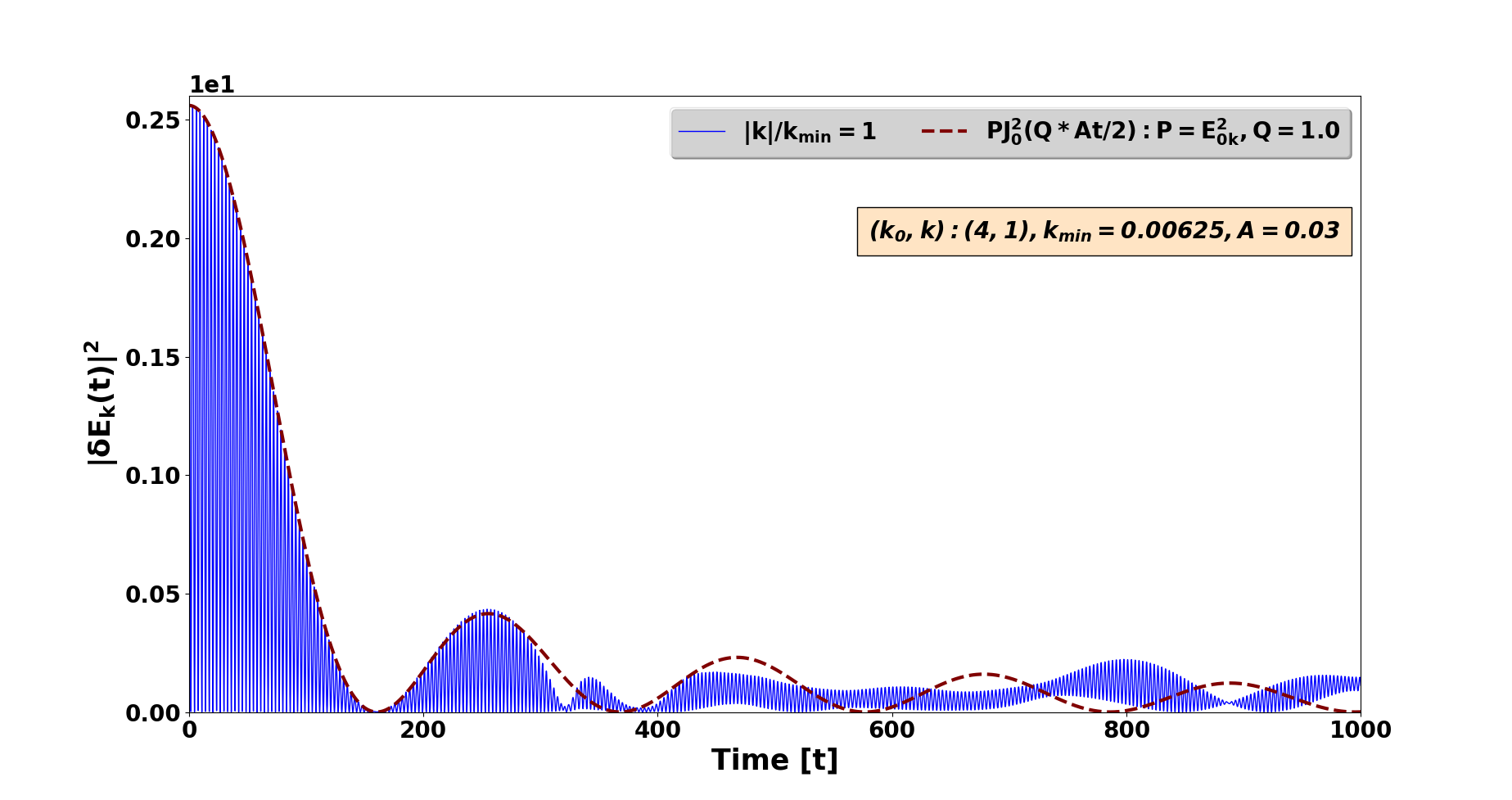}
\caption{ Portrait of temporal evolution of perturbed electric field mode $|\delta E_{k}|^{2}$ obtained from kinetic solver for $(k_{0},k):(4,1)$ pair, $k_{min}=0.00625,~A=0.03,~\alpha=0.01,~N_{x} \times N_{v} = [32768 \times 10000]$ alongwith zeroth order Bessel function $PJ_{0}^{2}(QAt/2)$ with $P=E_{0k}^{2}:E_{0k}=\alpha/k$ and $Q=1.0$. In agreement with Kaw et al. \cite{kaw1973}, our results demonstrate that the Bessel function correctly predicts the initial fall in the energy content of the perturbed primary mode $k=0.00625$ upto $t=300~\omega_{pe}^{-1}$ (approximately).}
\label{41_EK+J0_ALPHA=0.01_A=0.03_kmin=0.00625}
\end{figure}
\begin{figure*}
\centerline{\includegraphics[scale=0.37]{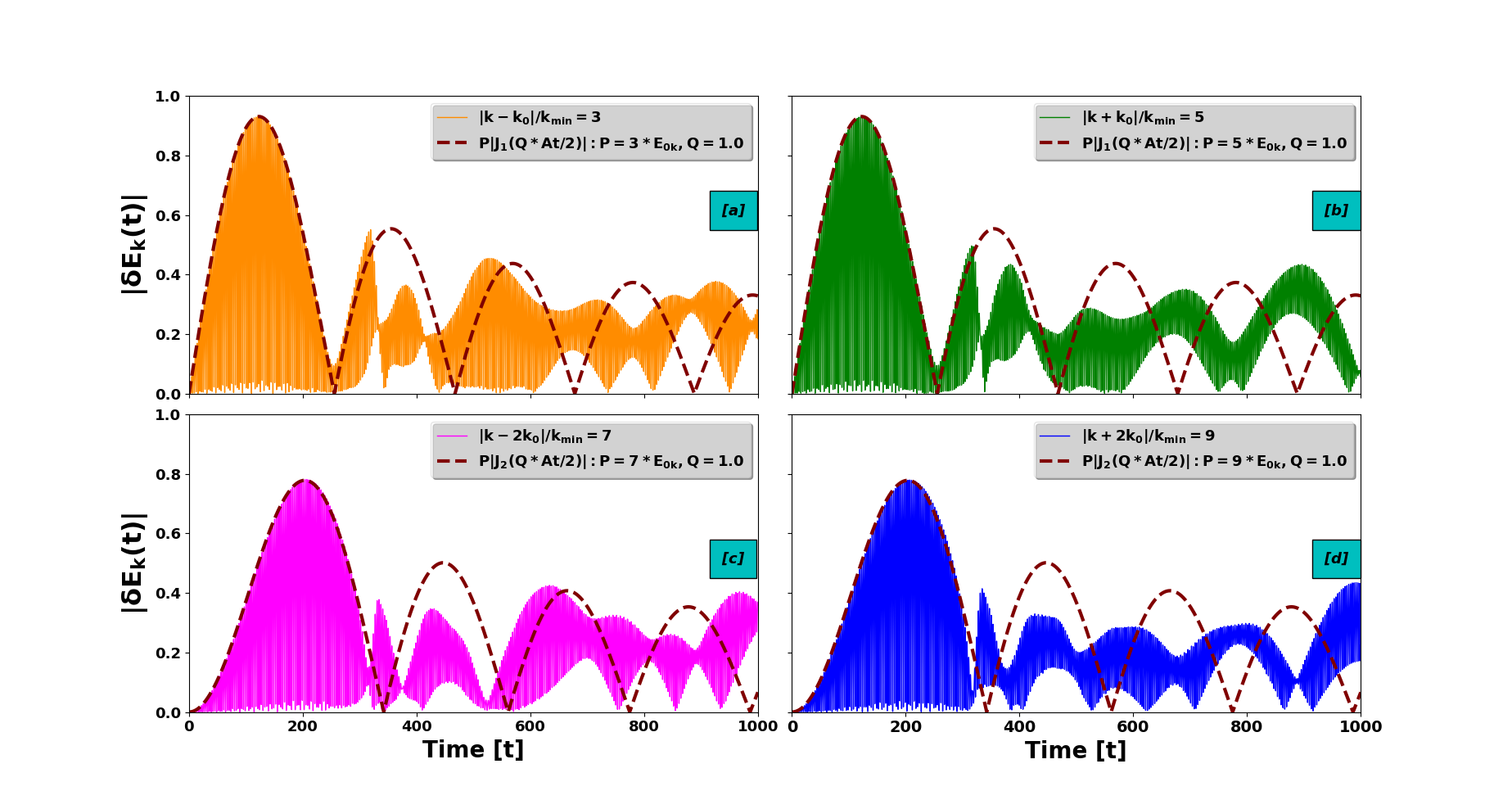}}
\caption{Temporal electric field signature $|\delta E_{k}|$ of coupled sideband modes $|k \pm Nk_{0}|:$N $\longrightarrow 1,~2$ i.e (a)$|k-k_{0}|/k_{min}$=3, (b)$|k + k_{0}|/k_{min}$=5, (c)$|k-2k_{0}|/k_{min}$=7 and (d)$|k+2k_{0}|/k_{min}$=9 obtained from kinetic solver for $(k_{0},k):(4,1)$ pair, $k_{min}=0.00625,~A=0.03,~\alpha=0.01,~N_{x} \times N_{v} = [32768 \times 10000]$ alongwith corresponding $1^{st}$ and $2^{nd}$ order Bessel functions i.e $|PJ_{1}(QAt/2)|$ and $|PJ_{2}(QAt/2)|$ respectively with $Q=1.0$.}
\label{41_EK+J0_3579_sidebands_A=0.03}
\end{figure*}
\begin{figure*}
\centerline{\includegraphics[scale=0.37]{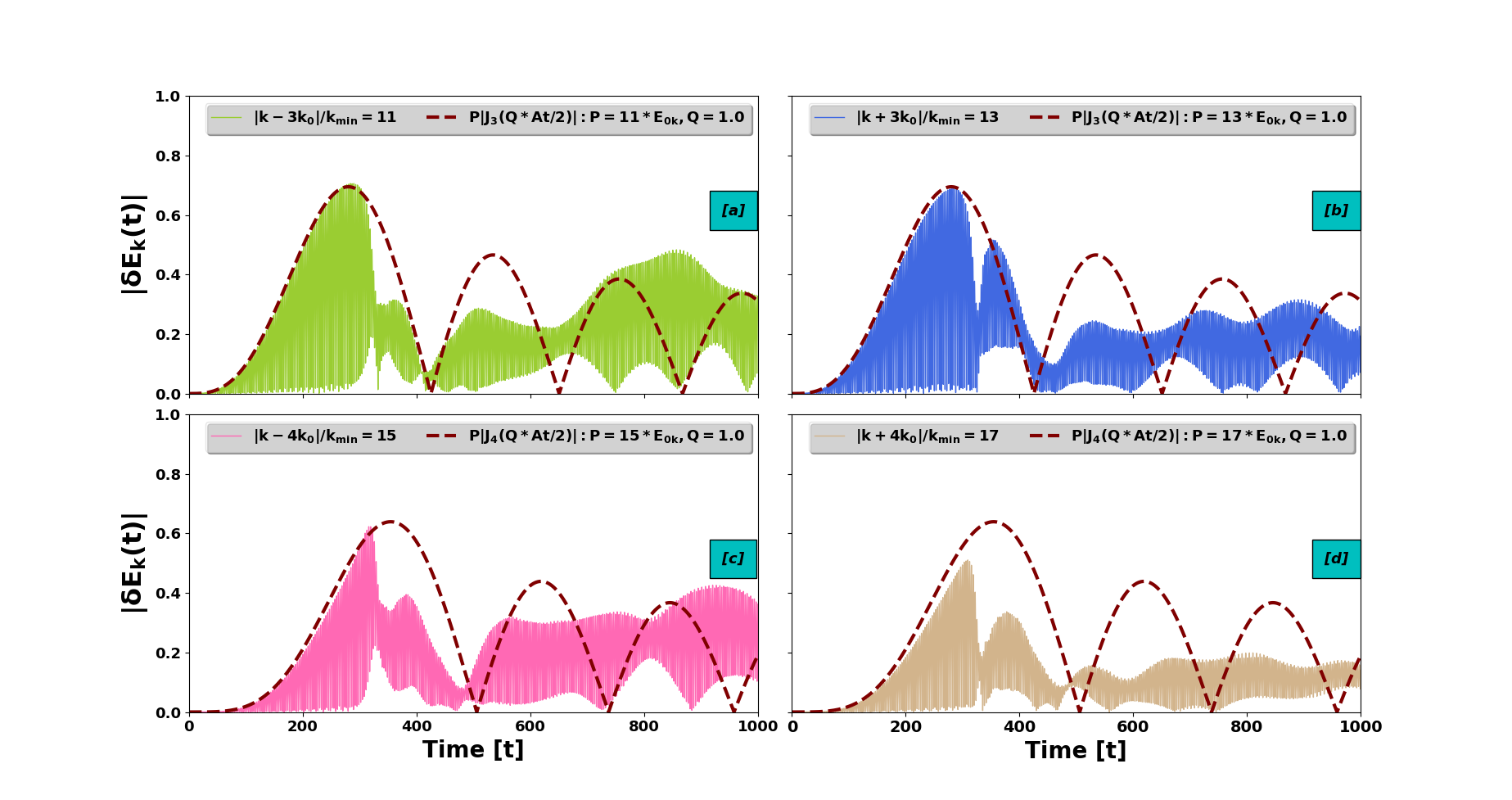}}
\caption{Temporal electric field signature $|\delta E_{k}|$ of coupled sideband modes $|k \pm Nk_{0}|:$N $\longrightarrow 3,4$ i.e (a)$|k-3k_{0}|/k_{min}$=11, (b)$|k + 3k_{0}|/k_{min}$=13, (c)$|k-4k_{0}|/k_{min}$=15 and (d)$|k+4k_{0}|/k_{min}$=17 obtained from kinetic solver for $(k_{0},k):(4,1)$ pair, $k_{min}=0.00625,~A=0.03,~\alpha=0.01,~N_{x} \times N_{v} = [32768 \times 10000]$ alongwith corresponding $3^{rd}$ and $4^{th}$ order Bessel functions i.e $|PJ_{3}(QAt/2)|$ and $|PJ_{4}(QAt/2)|$ respectively with $Q=1.0$.}
\label{41_EK+J0_11131517_sidebands_A=0.03}
\end{figure*}

   Figs. \ref{41_EK+J0_ALPHA=0.01_A=0.03_kmin=0.00625}, \ref{41_EK+J0_3579_sidebands_A=0.03} and \ref{41_EK+J0_11131517_sidebands_A=0.03} demonstrates the temporal evolution of perturbed $|E_{k}|$ as well as coupled $|E_{k \pm Nk_{0}}|$ modes obtained from kinetic solver for $(k_{0},k):(4,1)$ pair, $k_{min}=0.00625,~A=0.03,~\alpha=0.01,~N_{x} \times N_{v} = [32768 \times 10000]$ alongwith zeroth and $N^{th}$ order Bessel functions i.e $PJ_{0}^{2}(QAt/2)$ and $PJ_{N}(QAt/2)$ respectively with $P=E_{0k}^{2}~or~NE_{0k}:E_{0k}=\alpha/k$ and $Q=1.0$. From Fig. \ref{41_EK+J0_ALPHA=0.01_A=0.03_kmin=0.00625}, it is evident that the zeroth order Bessel function $PJ_{0}^{2}(QAt/2)$ where $P=E_{0k}^{2}:E_{0k}=\alpha/k=0.01/0.00625=1.60,~Q=1.0$ predicts exactly the initial fall in the energy density of the perturbed mode upto the peak of the second cycle i.e $t=300~\omega_{pe}^{-1}$ (approximately) as predicted by Kaw et al. \cite{kaw1973}. However, at late times the discrepancy away from the Bessel function scaling, could be either due to finite temperature effects or asynchronous energy exchange from perturbed to higher coupled sideband modes. Similarly from Fig. \ref{41_EK+J0_3579_sidebands_A=0.03} [(a), (b), (c) and (d)], it is clearly demonstrated that the temporal evolution of the Fourier amplitude of coupled sideband $|k \pm Nk_{0}|=3,5,7,9:N \longrightarrow 1,2$ modes can be exactly predicted via corresponding $N^{th}$ order Bessel function i.e  $PJ_{N}(QAt/2):N=1,2$ scaling where $P=NE_{0k},~Q=1.0$ upto the first cycle and the field amplitude signatures starts to deviate from the scaling after time $t>300~\omega_{pe}^{-1}$. Also, Fig. \ref{41_EK+J0_11131517_sidebands_A=0.03} [(a), (b), (c) and (d)] indicates that for the higher order coupled sideband $|k \pm Nk_{0}|=11,13,15,17:N \longrightarrow 3,4$ modes, the deviations of Bessel scaling from the corresponding Fourier amplitude $|E_{k \pm Nk_{0}}|$ becomes significant and the scaling only predicts half cycle amplitude $|E_{k \pm Nk_{0}}|$ signature. Hence to summarize, using our kinetic code VPPM-OMP 1.0 \cite{sanjeev2021}, we have clearly illustrated that temporal evolution of energy content of the Fourier amplitudes of perturbed and corresponding coupled sideband modes of a long-wavelength electron plasma wave (EPW) with $k=k_{min}=0.00625$ can be approximated using $N^{th}$ order Bessel function $J_{N}(x)$ scaling as predicted by Kaw et al. \cite{kaw1973} in their study using warm fluid model.

\begin{figure}
\centering{\includegraphics[scale=0.28]{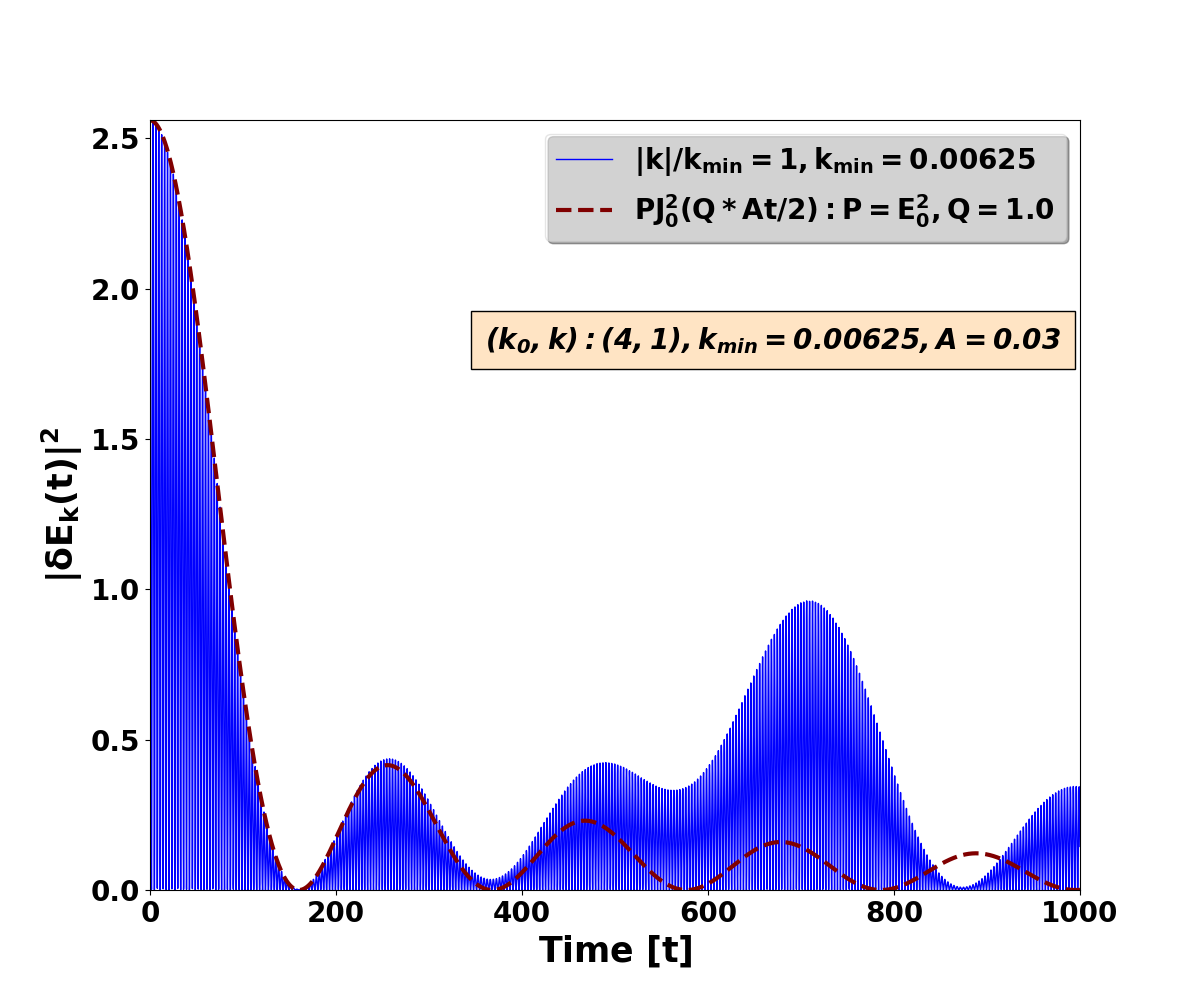}}
\caption{ Evolution of perturbed $|\delta E_{k}|^{2}$ mode with respect to time $(\omega_{pe}^{-1})$ obtained from warm fluid (BOUT++) solver for $(k_{0},k):(4,1)$ pair, $k_{min}=0.00625,~A=0.03,~\alpha=0.01$ alongwith zeroth order Bessel function $PJ_{0}^{2}(QAt/2)$ with $P=E_{0k}^{2}=2.56:E_{0k}=\alpha/k=1.6$ and $Q=1.0$.}
\label{FLUID_41_EK+J0_ALPHA=0.01_A=0.03_kmin=0.00625}
\end{figure}

\begin{figure*}
\centering{\includegraphics[height=10.5cm, width=15cm]{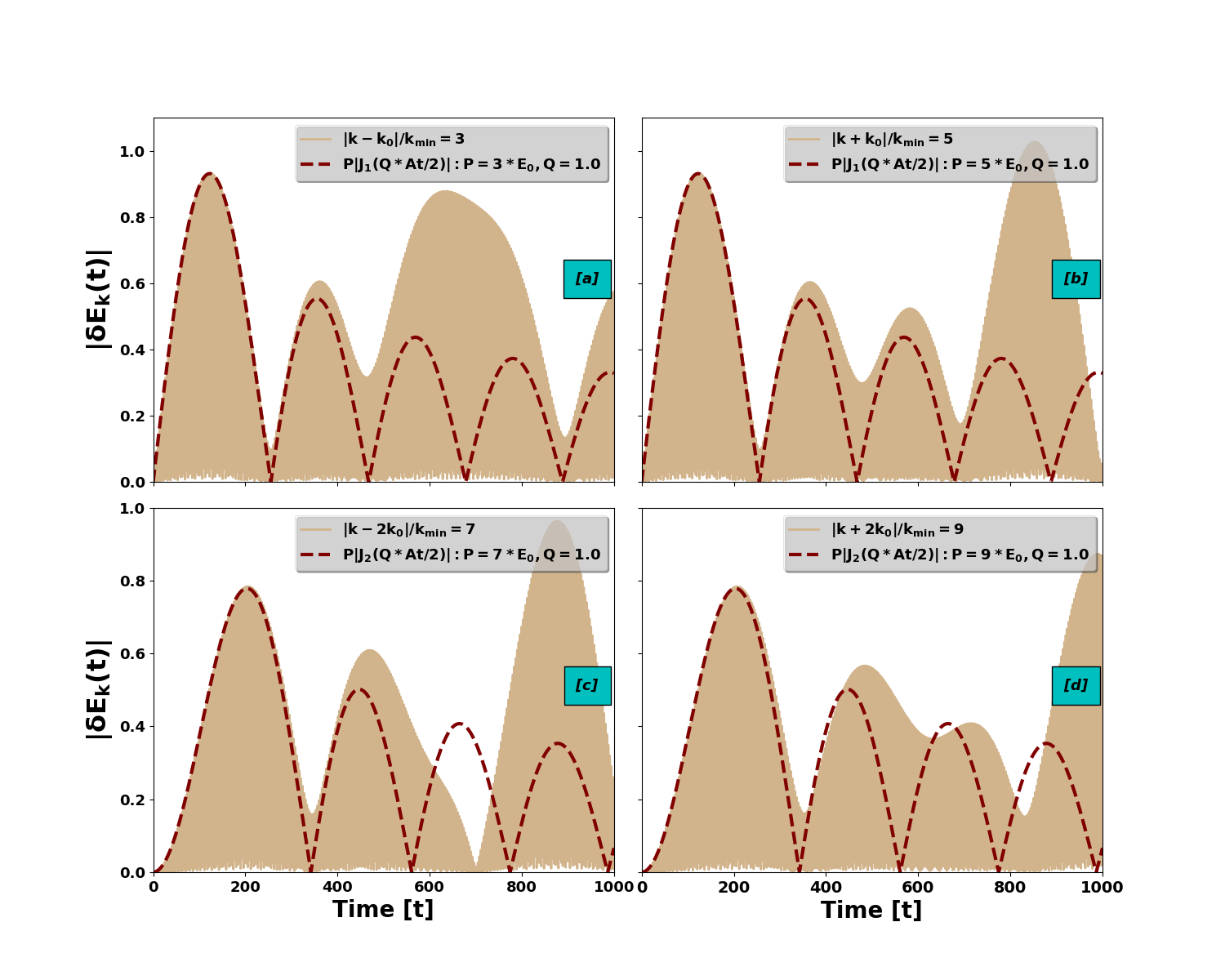}}
\caption{Variation of electric field signature $|\delta E_{k}|$ of coupled sideband modes $|k \pm Nk_{0}|:$N $\longrightarrow 1,~2$ i.e (a)$|k-k_{0}|/k_{min}$=3, (b)$|k + k_{0}|/k_{min}$=5, (c)$|k-2k_{0}|/k_{min}$=7 and (d)$|k+2k_{0}|/k_{min}$=9 with respect to time obtained from fluid (BOUT++) solver for $(k_{0},k):(4,1)$ pair, $k_{min}=0.00625,~A=0.03,~\alpha=0.01$ alongwith corresponding $1^{st}$ and $2^{nd}$ order Bessel functions i.e $|PJ_{1}(QAt/2)|$ and $|PJ_{2}(QAt/2)|$ respectively with $Q=1.0$.}
\label{FLUID_41_EK+J0_3579_sidebands_A=0.03}
\end{figure*}
\begin{figure*}
\centering{\includegraphics[height=11cm, width=15cm]{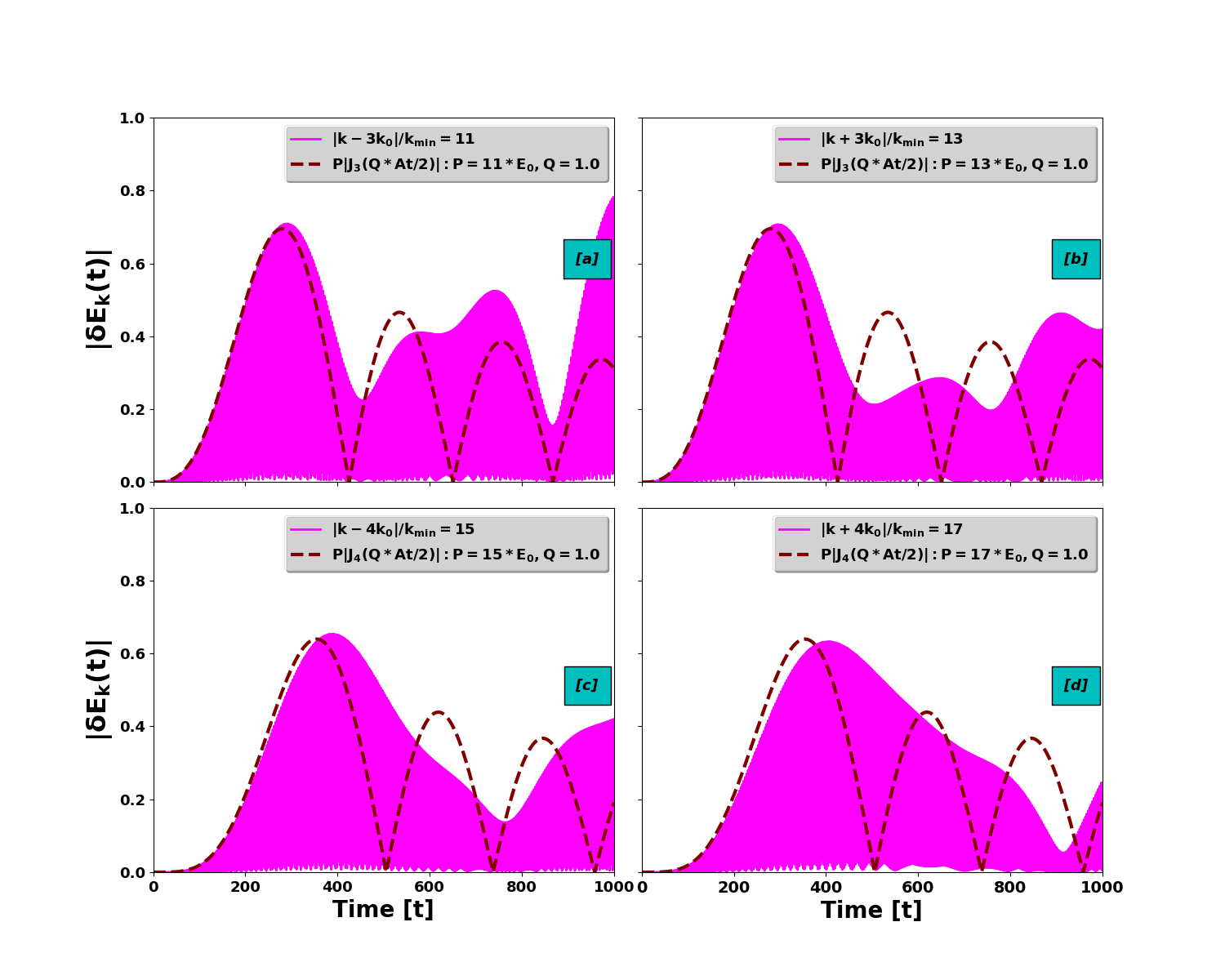}}
\caption{Variation of electric field signature $|\delta E_{k}|$ of coupled sideband modes $|k \pm Nk_{0}|:$N $\longrightarrow 3,~4$ i.e (a)$|k-3k_{0}|/k_{min}$=11, (b)$|k + 3k_{0}|/k_{min}$=13, (c)$|k-4k_{0}|/k_{min}$=15 and (d)$|k+4k_{0}|/k_{min}$=17 obtained from fluid (BOUT++) solver for $(k_{0},k):(4,1)$ pair, $k_{min}=0.00625,~A=0.03,~\alpha=0.01$ alongwith corresponding $3^{rd}$ and $4^{th}$ order Bessel functions i.e $|PJ_{3}(QAt/2)|$ and $|PJ_{4}(QAt/2)|$ respectively with $Q=1.0$. It is evident that for higher sideband modes, significant deviations from the Bessel scaling can be seen even before $t=360~\omega_{pe}^{-1}$.}
\label{FLUID_41_EK+J0_11131517_sidebands_A=0.03}
\end{figure*}

     To compare with the fluid simulation results, we have solved Eq. \ref{EQ_FINAL_5_FLUID} i.e warm fluid equation for plasma oscillations in the presence of background inhomogeneity derived from warm fluid model (mentioned in Secs. \ref{Fluid Model} and \ref{Numerical Scheme}), unlike the earlier theory \cite{kaw1973}, which assumes $k \longrightarrow 0$, Eq. \ref{EQ_FINAL_5_FLUID} is solved without any approximations using BOUT++ for exact parameter set as considered in the Vlasov simulations. Figs. \ref{FLUID_41_EK+J0_ALPHA=0.01_A=0.03_kmin=0.00625}, \ref{FLUID_41_EK+J0_3579_sidebands_A=0.03} and \ref{FLUID_41_EK+J0_11131517_sidebands_A=0.03} illustrates evolution of perturbed $|\delta E_{k}|^{2}$ and coupled sideband $|\delta E_{k \pm Nk_{0}}|$ modes obtained from fluid solver for $(k_{0},k):(4,1)$ pair, $k_{min}=0.00625,~A=0.03,~\alpha=0.01$ alongwith zeroth and $N^{th}$ order Bessel function scalings i.e $PJ_{0}^{2}(QAt/2)$ and $PJ_{N}(QAt/2)$ respectively with $P=E_{0k}^{2}~or~NE_{0k}:E_{0k}=\alpha/k$ and $Q=1.0$. Fig. \ref{FLUID_41_EK+J0_ALPHA=0.01_A=0.03_kmin=0.00625} shows that initial fall in the energy density of perturbed mode $|\delta E_{k}|$ which accurately follows the zeroth order Bessel function $J_{0}(x)$ scaling upto two cycles i.e till $t=360~\omega_{pe}^{-1}$ as predicted by Kaw et al. \cite{kaw1973}. However, it is evident that even the accurate numerical solutions obtained from a fluid solver, without any approximations, deviates from Bessel scaling at late times beyond $t>t_{0}$ where $t_{0}=360~\omega_{pe}^{-1}$ for this parameter set. Noticeable amount of deviation or increase in the amplitude of primary $|\delta E_{k}|^{2}$ can be seen in the fluid solutions around $t=600~\omega_{pe}^{-1}$ to $t=800~\omega_{pe}^{-1}$ range, which is absent when compared to the Vlasov solutions [See Figs. \ref{FLUID_41_EK+J0_ALPHA=0.01_A=0.03_kmin=0.00625} and \ref{41_EK+J0_ALPHA=0.01_A=0.03_kmin=0.00625}]. It may occur due to the finite temperature effects or kinetic effects which led to he damping of small fraction of energy density associated with the coupled sideband modes.

  Similarly, Fig. \ref{FLUID_41_EK+J0_3579_sidebands_A=0.03} [(a), (b), (c) and (d)] and Fig. \ref{FLUID_41_EK+J0_11131517_sidebands_A=0.03} [(a), (b), (c) and (d)] demonstrates Fourier mode amplitude evolution of coupled sideband modes $|\delta E_{k \pm Nk_{0}}|=3,5,7,9,11,13,15,17:N \longrightarrow 1-4$ with respect to time obtained from fluid (Bout++) solver for $(k_{0},k):(4,1)$ pair, $k_{min}=0.00625,~A=0.03,~\alpha=0.01$ case alongwith respective corresponding $N^{th}$ order Bessel functions i.e $|PJ_{N}(QAt/2)|$ where $P=NE_{0k},~Q=1.0$. For adjacent sideband modes with $N=1,2$, Fig. \ref{FLUID_41_EK+J0_3579_sidebands_A=0.03} [(a), (b), (c) and (d)] indicates that the Bessel scaling only predicts one and half cycle of the temporal evolution of electric field $|\delta E_{k \pm Nk_{0}}|$ fluid solution until $t=360~\omega_{pe}^{-1}$ and strong divergence with surge in $|\delta E_{k \pm Nk_{0}}|$ solutions can be seen at late time after $t>360~\omega_{pe}^{-1}$. Meanwhile, for coupled sideband modes with $N=3,4$, the deviation from the respective Bessel scaling occurs earlier in time and are more prominent. In comparison with the solution obtained from kinetic solver, the above discussed results are qualitatively similar to the solutions obtained from fluid solver but the Fourier amplitude signature of these coupled sideband modes $|\delta E_{k \pm Nk_{0}}|$ are significantly different due to the absence of the involved kinetic effects not addressed in the fluid model. In the next section \ref{Finite inhomogeneity amplitude effects}, we will present the effect of ion inhomogeneity amplitude $A$ on the evolution of perturbed long-wavelength EPW mode alongwith comparative studies between fluid and kinetic solutions.


\subsection{Finite inhomogeneity amplitude effects}
\label{Finite inhomogeneity amplitude effects}
\begin{figure*}
\centering{\includegraphics[height=11cm, width=15cm]{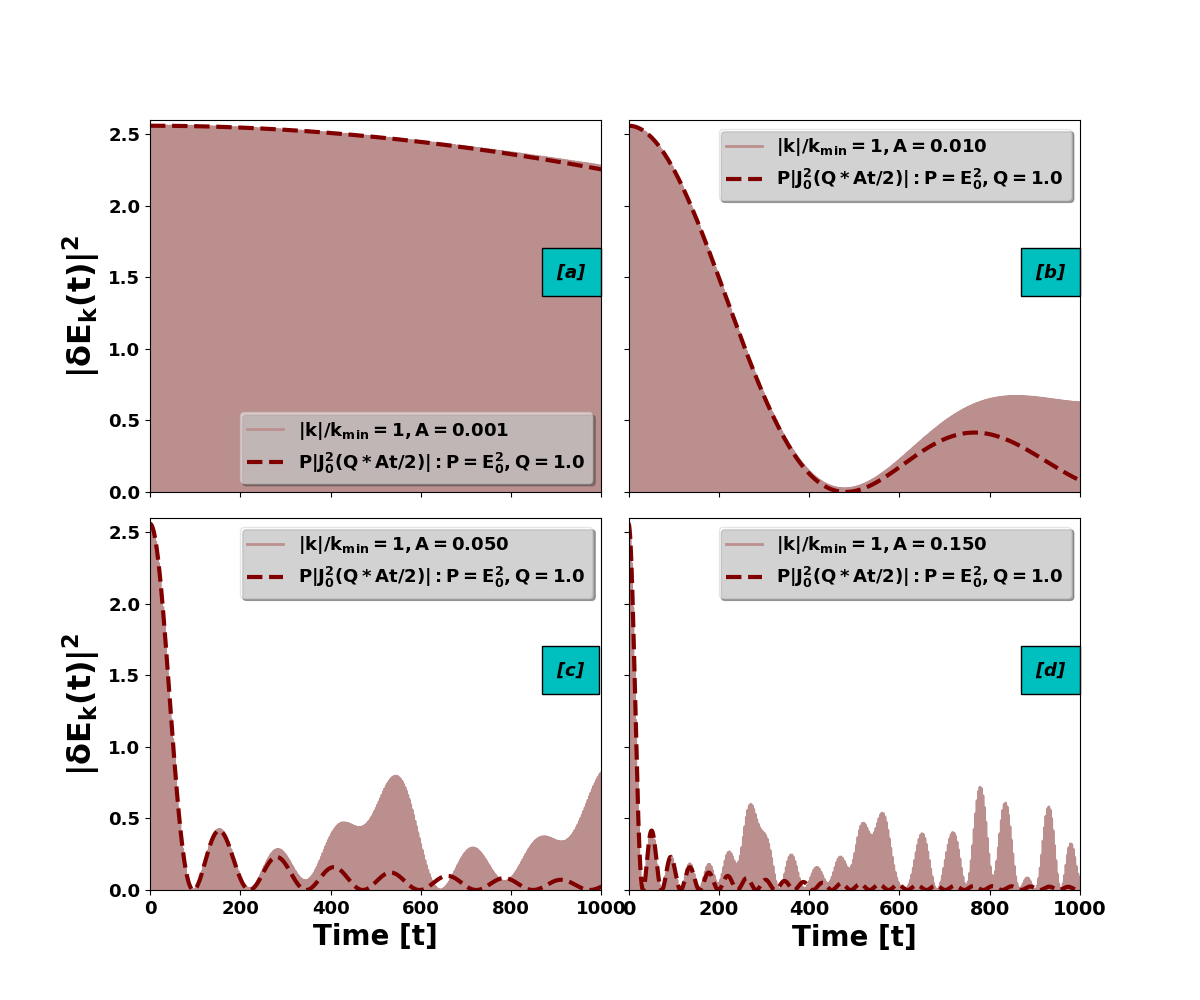}}
\caption{Perturbation Fourier mode $|\delta E_{k}|^{2}$ signature versus time $(t)$ obtained from fluid (BOUT++) solver for $(k_{0},k):(4,1):k_{min}=0.00625$ with inhomogeneity amplitudes $A$ viz (a) $A=0.001$, (b) $A=0.01$, (c) $A=0.05$ and (d) $A=0.15$. Also, zeroth order Bessel function $PJ_{0}^{2}(QAt/2)$ with $P=E_{0k}^{2}:E_{0k}=\alpha/k$ and $Q=1.0$ is also plotted alongside each cases. It shows that for small values of $A$ the Bessel scaling exactly matches the $|\delta E_{k}|$ evolution whereas when the $A$ value is increased the divergence between the two occurs at early times. }
\label{FLUID_41_EK2+JO2_ALPHA=0.001+0.01+0.05+0.15_SUBPLOT}
\end{figure*}

\begin{figure*}
\centerline{\includegraphics[scale=0.38]{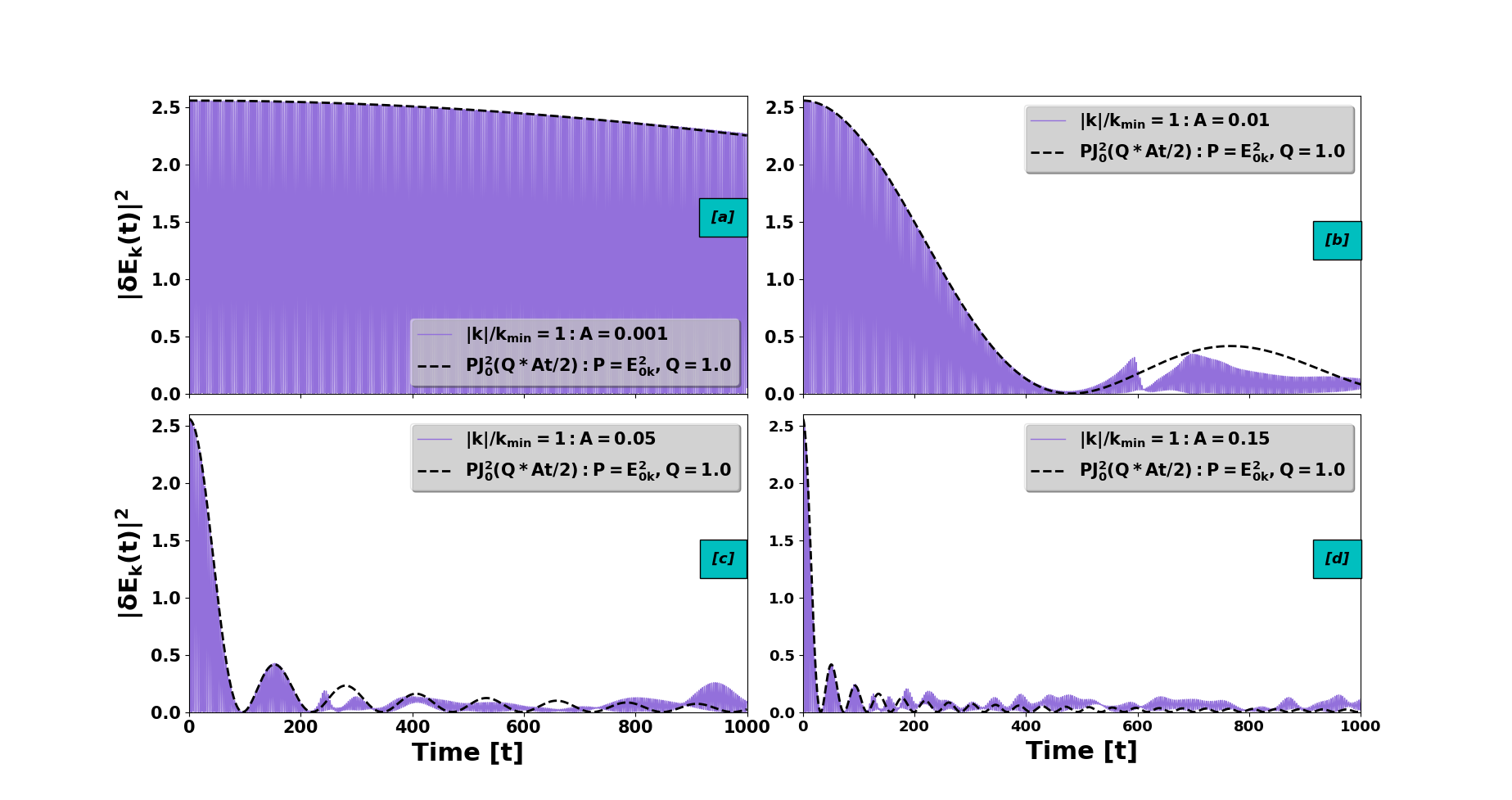}}
\caption{Evolution of perturbation Fourier amplitudes $|\delta E_{k}|^{2}$ versus time $(t)$ for $(k_{0},k):(4,1):k_{min}=0.006225$ obtained from kinetic solver with various inhomogeneity amplitudes $A$ i.e (a) $A=0.001$, (b) $A=0.01$, (c) $A=0.05$ and (d) $A=0.15$. Also, zeroth order Bessel function $PJ_{0}^{2}(QAt/2)$ with $P=E_{0k}^{2}:E_{0k}=\alpha/k$ and $Q=1.0$ is also plotted alongside each cases. It is evident that for small values of $A$ the Bessel function scaling exactly matches the perturbation amplitude evolution whereas when the $A$ value is increased the deviation between the two occurs very early in time.   }
\label{41_EK2+JO2_ALPHA=0.001+0.01+0.05+0.15_SUBPLOT}
\end{figure*}

   In this Section, we will illustrate the effect of increase in inhomogeneity amplitude $(A)$ on the temporal evolution of long-wavelength EPW mode for a given (i.e, fixed) amplitude value $\alpha$ of perturbation $k$. With rest of the simulation parameters initialized identical as before for linear studies in both the kinetic and fluid solvers i.e $k_{min}=0.00625,~(k_{0},k):(4,1),~\alpha=0.01$, we investigate the evolution of the perturbation Fourier amplitudes $|\delta E_{k}|^{2}$ solutions obtained from fluid and kinetic solvers versus time for a range of $A$ values from 0.001 to 0.15 alongwith zeroth order Bessel function $PJ_{0}^{2}(QAt/2)$ scaling as shown in Fig. \ref{FLUID_41_EK2+JO2_ALPHA=0.001+0.01+0.05+0.15_SUBPLOT} [(a), (b), (c), and (d)] and Fig. \ref{41_EK2+JO2_ALPHA=0.001+0.01+0.05+0.15_SUBPLOT} [(a), (b), (c), and (d)] respectively.

  It is apparent from the Fig. \ref{FLUID_41_EK2+JO2_ALPHA=0.001+0.01+0.05+0.15_SUBPLOT} (a) and Fig \ref{41_EK2+JO2_ALPHA=0.001+0.01+0.05+0.15_SUBPLOT} (a) that for small $A$ values the zeroth  order Bessel function $J_{0}^{2}(At/2)$ accurately scales the amplitude evolution $|\delta E_{k}|^{2}$ till the end of the simulation time i.e $t=1000 ~ \omega_{pe}^{-1}$ whereas when we increase the inhomogeneity amplitude $A$, Bessel scaling $J_{0}^{2}(At/2)$ starts to deviate from the field signature earlier in time. Also, from Figs. \ref{FLUID_41_EK2+JO2_ALPHA=0.001+0.01+0.05+0.15_SUBPLOT} [(b), (c) and (d)] and \ref{41_EK2+JO2_ALPHA=0.001+0.01+0.05+0.15_SUBPLOT} [(b), (c) and (d)], it is quite clear that for large amplitude inhomogeneity parameter sets the zeroth order Bessel function $J_{0}^{2}(At/2)$ scaling estimates more cycles of amplitude evolution compared to the small inhomogeneity amplitudes. Due to the absence of kinetic effects in solutions obtained from the fluid solver, we can see noticeable electric field amplitude peaks at late times in Fig. \ref{FLUID_41_EK2+JO2_ALPHA=0.001+0.01+0.05+0.15_SUBPLOT} [(b), (c) and (d)] when compared to the kinetic solutions in Fig. \ref{41_EK2+JO2_ALPHA=0.001+0.01+0.05+0.15_SUBPLOT} [(b), (c) and (d)]. 

\begin{figure*}
\centerline{\includegraphics[scale=0.38]{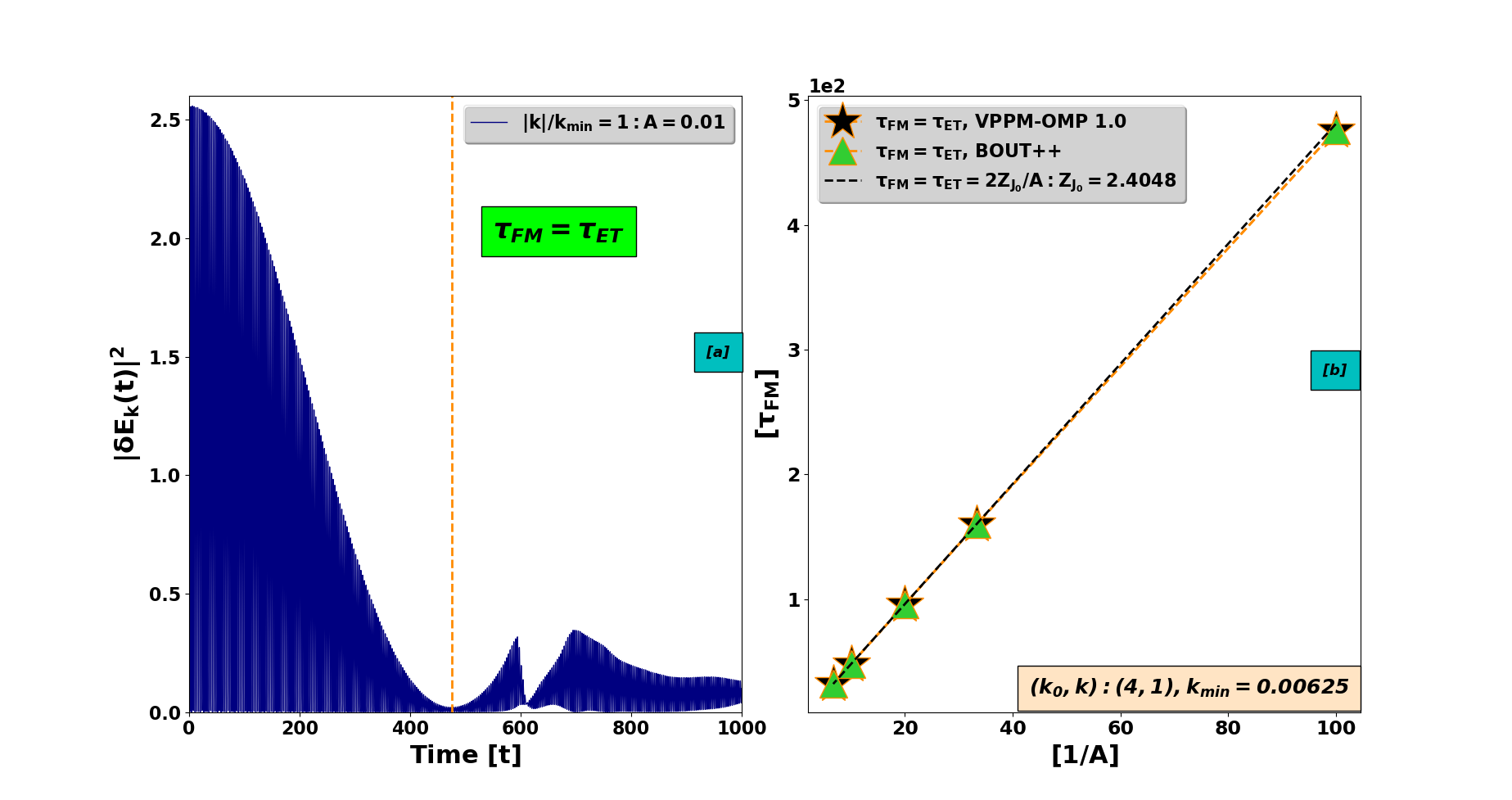}}
\caption{Portrait of (a) perturbation $k/k_{min}=1$ Fourier amplitude $|\delta E_{k}|^{2}$ versus time $(t)$ obtained using kinetic solver for $(k_{0},k):(4,1):k_{min}=0.00625$ with $A=0.01$ and (b) Time taken $\tau_{FM}$ by the mode to reach the first minima of the energy density signature $|\delta E_{k}|^{2}$ versus inhomogeneity amplitude $A$ ranging from 0.01 to 0.15 with $(k_{0},k):(4,1):k_{min}=0.00625$ obtained using both kinetic (VPPM-OMP 1.0) and warm fluid (BOUT++) solvers as tabulated in Tables \ref{TABLE 2} and \ref{TABLE 3} respectively. It is clear from (b) that $\tau_{FM}$ when all the energy of the primary mode is coupled at $\tau_{FM}$, then $\tau_{FM}=\tau_{ET}$ (also called phase mixing time) is inversely proportional to inhomogeneity amplitude $A$ and scales as $\tau_{FM}=\tau_{ET}=2Z_{J_{0}}/A:Z_{J_{0}}= 2.4048$.}
\label{41_K0K_TAU_ET_VS_A_KMIN=0_00625}
\end{figure*}

\begin{table}
\caption{Time required to attain first minimum of the energy density signature i.e $\tau_{FM}~(\omega_{pe}^{-1})$ obtained for various $k_{min}$ and inhomogeneity amplitude $A$ values with $(k_{0},k):(4,1)$ and $\alpha=0.01$ using VPPM-OMP 1.0 kinetic solver alongwith energy transfer time given by Kaw et al. \cite{kaw1973} studies (Eq. \ref{EQ_TAU_VS_A}). Here, $Z_{J_{0}}= 2.4048$ is the first zero of $J_{0}(x)$ and $k=k_{min}$ for all the values obtained from kinetic solver.}  
\centering                         
\begin{tabular}{c c c c c c}           
\hline\hline                        
Inhomogeneity & Kaw et al. \cite{kaw1973} & [Kinetic] & [Kinetic] & [Kinetic] & [Kinetic] \\
Amplitude & $\tau_{FM}=(2Z_{J_{0}}/A)$ & $\tau_{FM}$ & $\tau_{FM}$ & $\tau_{FM}$ & $\tau_{FM}$ \\
A & $k \rightarrow 0$ & $k_{min}$ & $k_{min}$ & $k_{min}$ & $k_{min}$ \\
$~$ & $~$ & $0.00625$ & $0.0125$ & $0.025$ & $0.1$ \\ [1.0ex]    
    
\hline            
0.01 & 480.96 & 476.0 & 431.0 & 361.0 & 2569.0  \\          
0.03 & 160.32 & 160.0 & 159.0 & 140.0 & 1099.0 \\
0.05 & 96.192 & 96.0 & 96.0 & 89.0 & 645.0 \\
0.10 & 48.096 & 48.0 & 46.0 & 49.0 & 290.0 \\
0.15 & 32.064 & 32.0 & 32.0 & 30.0 & 207.0 \\ [1ex]
\hline                              
\end{tabular}
\label{TABLE 2}
\end{table}

\begin{table}
\caption{Time required to attain first minimum of the energy density signature i.e $\tau_{FM}~(\omega_{pe}^{-1})$ obtained for various $k_{min}$ and inhomogeneity amplitude $A$ values with $(k_{0},k):(4,1)$ and $\alpha=0.01$ using BOUT++ warm fluid solver alongwith energy transfer time given by \cite{kaw1973} studies (Eq. \ref{EQ_TAU_VS_A}). Here, $Z_{J_{0}}= 2.4048$ is the first zero of $J_{0}(x)$ and $k=k_{min}$ for all the values obtained from fluid solver.}  
\centering                         
\begin{tabular}{c c c c c c}           
\hline\hline                        
Inhomogeneity & Kaw et al. \cite{kaw1973} & [Fluid] & [Fluid] & [Fluid] & [Fluid] \\
Amplitude & $\tau_{FM}=(2Z_{J_{0}}/A)$ & $\tau_{FM}$ & $\tau_{FM}$ & $\tau_{FM}$ & $\tau_{FM}$ \\
A & $k \rightarrow 0$ & $k_{min}$ & $k_{min}$ & $k_{min}$ & $k_{min}$ \\    
$~$ & $~$ & $0.00625$ & $0.0125$ & $0.025$ & $0.1$ \\ [1.0ex]    
\hline            
0.01 & 480.96 & 476.0 & 431.0 & 370.7 & 34.0  \\          
0.03 & 160.32 & 160.0 & 159.0 & 138.9 & 28.0 \\
0.05 & 96.192 & 96.0 & 96.0 & 88.9 & 27.9 \\
0.10 & 48.096 & 48.0 & 48.0 & 48.8 & 28.0 \\
0.15 & 32.064 & 32.0 & 32.0 & 31.0 & 25.124 \\ [1ex]
\hline                               
\end{tabular}
\label{TABLE 3}
\end{table}

  Let us now dvelve on the time taken $\tau_{FM}$ by the mode to reach the first minima of the energy density signature $|\delta E_{k}|^{2}$. Also, for small perturbation scale $k$ such as $k=0.00625$ or $k \longrightarrow 0$, $\tau_{FM}$ is energy transfer time $\tau_{ET}$ (sometimes also called as phase mixing time) \cite{kaw1973}. It is the time at which most of the energy content in a primary perturbation mode is transfered to the coupled sideband modes as shown in Figs. \ref{41_EK+J0_ALPHA=0.01_A=0.03_kmin=0.00625}, \ref{41_EK+J0_3579_sidebands_A=0.03} and \ref{41_EK+J0_11131517_sidebands_A=0.03}. In Fig. \ref{41_K0K_TAU_ET_VS_A_KMIN=0_00625} [(a) and (b)], we  demonstrate the variation of perturbation Fourier amplitude $|\delta E_{k}|^{2}$ versus time obtained from kinetic and fluid solvers for $(k_{0},k):(4,1),~k_{min}=0.00625$ with $A=0.03$ and energy transfer time $\tau_{ET}$ versus inhomogeneity amplitude $A$ ranging from 0.01 to 0.15 obtained using both kinetic and fluid solvers alongwith energy transfer time estimated by theoretical studies of Kaw et al. \cite{kaw1973} are tabulated in Tables \ref{TABLE 2} and \ref{TABLE 3}. It is important to note that the energy transfer time $\tau_{ET}$ or $\tau_{FM}$ for solutions obtained from both kinetic and fluid solvers are exact and are in close estimation to $\tau_{ET}$ values given by Kaw et al. \cite{kaw1973}. Hence, we have traced the variation of $\tau_{ET}$ with respect to inhomogeneity amplitude $A$ which is true for both kinetic and fluid solutions (shown in Fig. \ref{41_K0K_TAU_ET_VS_A_KMIN=0_00625} (b)) and found out the empirical scaling law for $k_{min}=0.00625$ as,
 
\begin{equation}
 \tau_{FM}=\tau_{ET} = \left[ \frac{2Z_{J_{0}}}{A} \right] ~:~Z_{J_{0}}= 2.4048
\label{EQ_TAU_VS_A}
\end{equation}  
i.e in case of small perturbation scale $k$, most of the energy in perturbed $|\delta E_{k}|$ mode will get transfered quickly to coupled sideband $|E_{k \pm Nk_{0}}|$ modes for large inhomogeneity amplitude $A$ values due to mode coupling phenomenon to larger number of modes as the coupling parameter $N \propto \sqrt{A}$ (See Eq. \ref{EQ_N}) for a fixed background ion inhomogeneity scale $k_{0}$. Scaling law Eq. \ref{EQ_TAU_VS_A} is in agreement with Kaw et al. \cite{kaw1973} studies which reports the inverse scaling between $\tau_{ET}$ and $A$ as $\tau_{ET}=2Z_{J_{0}}/A:Z_{J_{0}}= 2.4048$.

\begin{figure*}
\centerline{\includegraphics[scale=0.38]{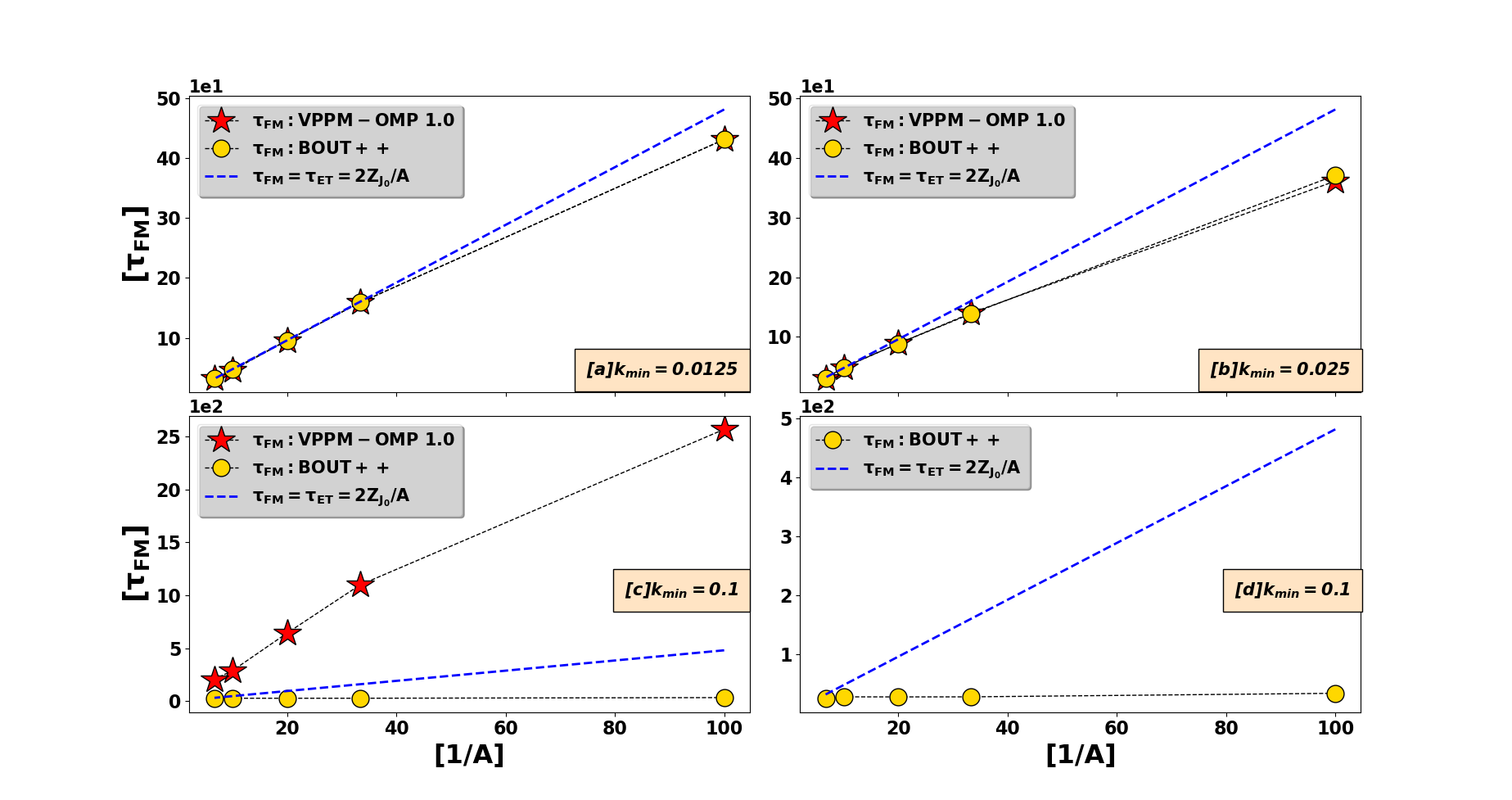}}
\caption{Variation of time required to attain first minimum of energy density signature i.e $\tau_{FM}$ with respect to inhomogeneity amplitude $A$ obtained using both kinetic (VPPM-OMP 1.0) and warm fluid (BOUT++) solvers for $\alpha=0.01,~(k_{0},k):(4,1)$ pair with different $k_{min}$ values i.e (a)$k_{min}=0.0125$, (b)$k_{min}=0.025$ and (c)$k_{min}=0.1$, tabulated in Tables \ref{TABLE 2} and \ref{TABLE 3}. In (d) zoomed plot on y-scale of (c) is shown. It is evident from (a), (b), (c) and (d) that the inverse scaling law given in Eq. \ref{EQ_TAU_VS_A} ($\tau_{FM}=\tau_{ET}=2Z_{J_{0}}/A$) does not hold true for all $k_{min}$ values eventhough for each case, $v_{\phi}(=\omega/k) \gg v_{thermal}$.}
\label{41_K0K_TAU_ET_VS_A_FOR_ALL_KMIN}
\end{figure*}
\begin{figure*}
\centerline{\includegraphics[scale=0.38]{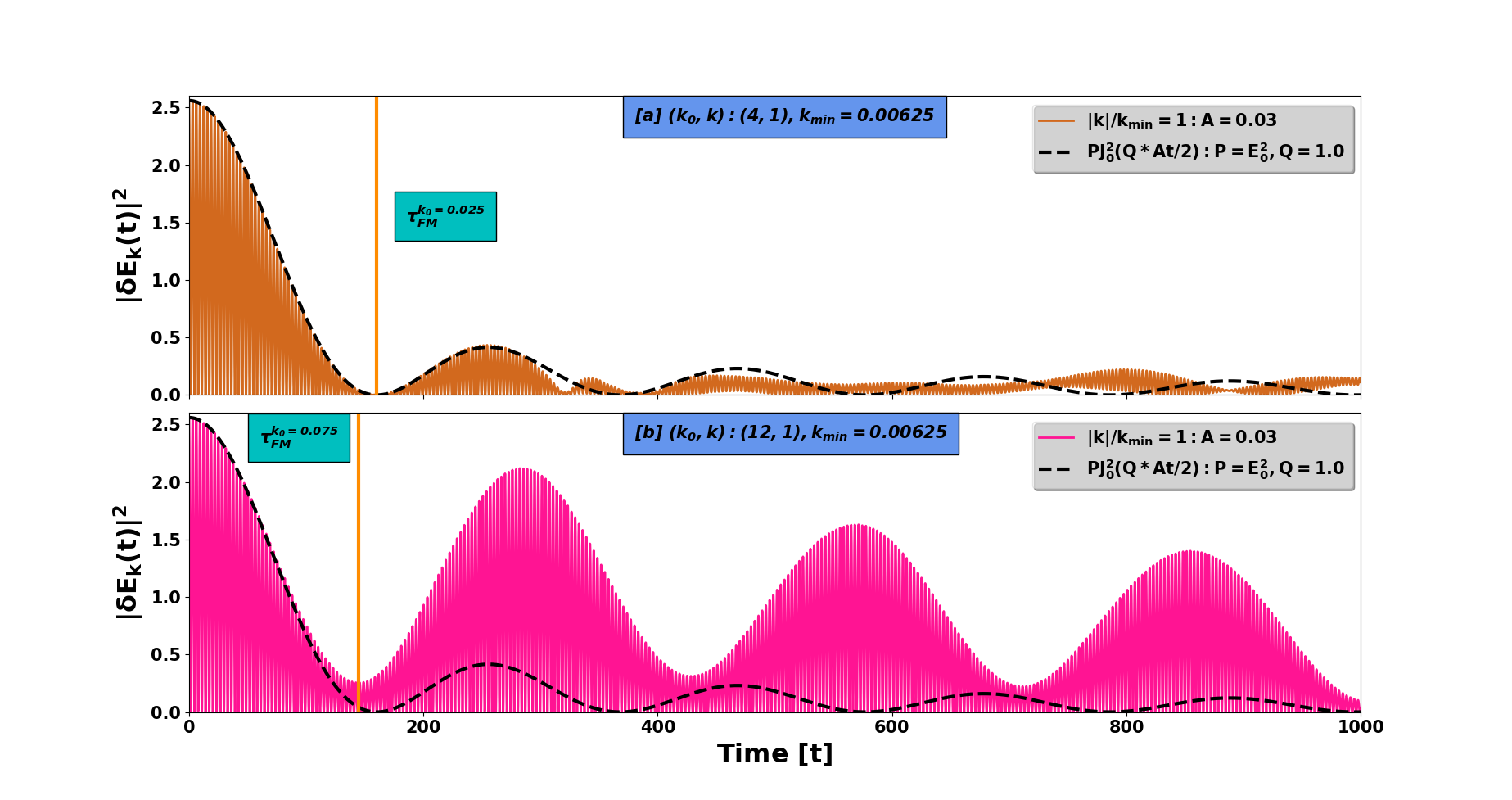}}
\caption{Time evolution of perturbed electric field $|\delta E_{k}|^{2}$ modes obtained from kinetic solver with $A=0.03,~\alpha=0.01,~k_{min}=0.00625,~N_{x} \times N_{v} = [32768 \times 10000]$ for (a) $(k_{0},k):(4,1)$ and (b) $(k_{0},k):(12,1)$ pairs. Time required to attain first minimum of energy density signature $|\delta E_{k}|^{2}$ i.e $\tau_{FM}$ for both cases are (a) $\tau_{FM}^{k_{0}=0.025}=160 ~\omega_{pe}^{-1}$ and (b) $\tau_{FM}^{k_{0}=0.075}=150 ~\omega_{pe}^{-1}$ as indicated by the solid lines. It portrays the change in $\tau_{FM}$ with the increase in the inhomogeneity scale $k_{0}$.}
\label{41+121_K0K_EK2+J02_A=0.03_KMIN=0.00625}
\end{figure*}

   We have also obtained the change of $\tau_{FM}$ versus $A$ with $\alpha=0.01,~(k_{0},k):(4,1)$ for a range of $A$ values ranging from 0.01 to 0.15 with different sets of $k_{min}$ values i.e (a)$k_{min}=0.0125$, (b)$k_{min}=0.025$ and (c)$k_{min}=0.1$ obtained using both kinetic (VPPM-OMP 1.0) and warm fluid (BOUT++) solvers as shown in Fig. \ref{41_K0K_TAU_ET_VS_A_FOR_ALL_KMIN} [(a), (b), (c) and (d)]. In Fig. \ref{41_K0K_TAU_ET_VS_A_FOR_ALL_KMIN} (d), zoomed plot on y-scale of (c) is shown which indicates the deviation between inverse scaling law given in Eq. \ref{EQ_TAU_VS_A} and the fluid solutions. Also, in Table \ref{TABLE 2} and \ref{TABLE 3}, $\tau_{FM}$ values are tabulated for various $k_{min}$ which are obtained using both kinetic (VPPM-OMP 1.0) and warm fluid (BOUT++) solvers. The empirical law given in Eq. \ref{EQ_TAU_VS_A} i.e the theoretical estimate of Kaw et al. \cite{kaw1973} (for $k \longrightarrow 0$), does not hold true for all finite $k_{min}$ values and it especially violates in the $k_{min}=0.1$ case which indicates that the $\tau_{FM}$ is also dependent on inhomogeneity scale $k_{0}$ alongwith inhomogeneity amplitude $A$. Also, from Fig. \ref{41_K0K_TAU_ET_VS_A_FOR_ALL_KMIN} [(b) and (c)], it is clear that the $\tau_{FM}$ values obtained from fluid solutions starts to strongly diverges from kinetic and linear (Eq. \ref{EQ_TAU_VS_A}) scalings with the gradual increase in $k_{min}$ values indicating the finite system size and kinetic effects on the time evolution of these EPW modes.
 
  In order to confirm the dependency of $\tau_{FM}$ on inhomogeneity scale $k_{0}$, we have calculated the time required to attain first minimum of energy density signature i.e $\tau_{FM}$ for two different pairs of $(k_{0},k)$ scales i.e $(4,1)$ and $(12,1)$ which implies $k_{0}=0.025$ and $k_{0}=0.075$ respectively with $k_{min}=0.00625$ for both cases. Fig. \ref{41+121_K0K_EK2+J02_A=0.03_KMIN=0.00625} [(a) and (b)] shows the variation of perturbed electric field $|\delta E_{k}|^{2}$ modes obtained from kinetic solver with $A=0.03,~\alpha=0.01,~k_{min}=0.00625,~N_{x} \times N_{v} = [32768 \times 10000]$ for (a) $(k_{0},k):(4,1)$ and (b) $(k_{0},k):(12,1)$ pairs alongwith Bessel $J_{0}^{2}(x)$ scaling. It is evident from Fig. \ref{41+121_K0K_EK2+J02_A=0.03_KMIN=0.00625} [(a) and (b)] that there is a shift in the $\tau_{FM}$ for both values of $k_{0}$ which is $\Delta \tau_{FM}= |\tau_{FM}^{k_{0}=0.025}-\tau_{FM}^{k_{0}=0.075}|=10~\omega_{pe}^{-1}: \tau_{FM}^{k_{0}=0.025}=160 ~\omega_{pe}^{-1},~ \tau_{FM}^{k_{0}=0.075}=150 ~\omega_{pe}^{-1}$. Hence we can conclude that $\tau_{FM}$ is dependent on both the inhomogeneity amplitude $A$ as well as inhomogeneity scale $k_{0}$, contrary to the Kaw et al. \cite{kaw1973} studies using warm fluid theory which reports that $\tau_{FM} \sim \tau_{ET}: k \longrightarrow0$ is independent of ion inhomogeneity scale $k_{0}$. Also, $N^{2} \propto 1/k_{0}^{2}$, hence for large $k_{0}$, $N \longrightarrow$ small implying primary mode energy will keep coming back as shown in Fig. \ref{41+121_K0K_EK2+J02_A=0.03_KMIN=0.00625} (b). So, this is in line with Kaw et al. \cite{kaw1973} model. In the following section \ref{Finite Size effects}, we present the discussions on effects of finite system size on the Bessel function $J_{N}(At/2)$ scaling and temporal evolution of perturbed long-wavelength EPW mode with different sets of $k_{min}$ values. 

\subsection{Finite Size effects}
\label{Finite Size effects}

\begin{figure*}
\centering{\includegraphics[height=11cm, width=15cm]{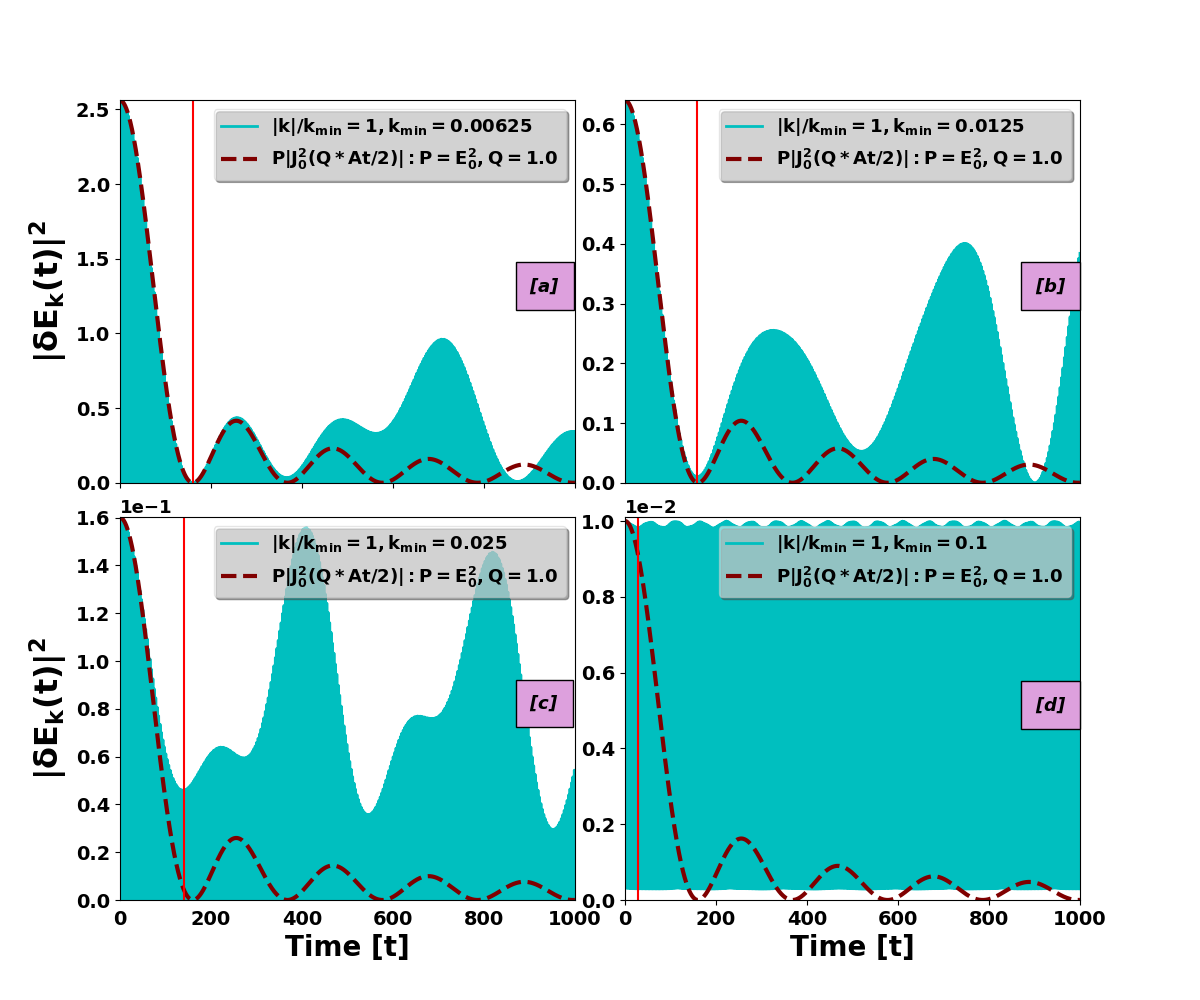}}
\caption{Temporal evolution of perturbation Fourier amplitudes $|\delta E_{k}|^{2}$ versus time $(t)$ obtained using fluid (BOUT++) solver for $(k_{0},k):(4,1),~\alpha=0.01,~A=0.03$ with various $k_{min}$ values i.e (a) $k_{min}=0.00625$, (b) $k_{min}=0.0125$, (c) $k_{min}=0.025$ and (d) $k_{min}=0.1$. Also, zeroth order Bessel function $PJ_{0}^{2}(QAt/2)$ with $P=E_{0k}^{2}:E_{0k}=\alpha/k$ and $Q=1.0$ is also plotted alongside each cases. It is evident from (a), (b), (c) and (d) that with the increase in the value of $k_{min}$ from 0.00625 to 0.1, Bessel $J_{0}^{2}(x)$ scaling strongly deviates from the perturbed Fourier amplitude $|\delta E_{k}|^{2}$.}
\label{FLUID_41_K0K_EK2+J02_A=0.03_KMIN=0.00625+0.0125+0.025+0.1}
\end{figure*}

\begin{figure*}
\centerline{\includegraphics[scale=0.38]{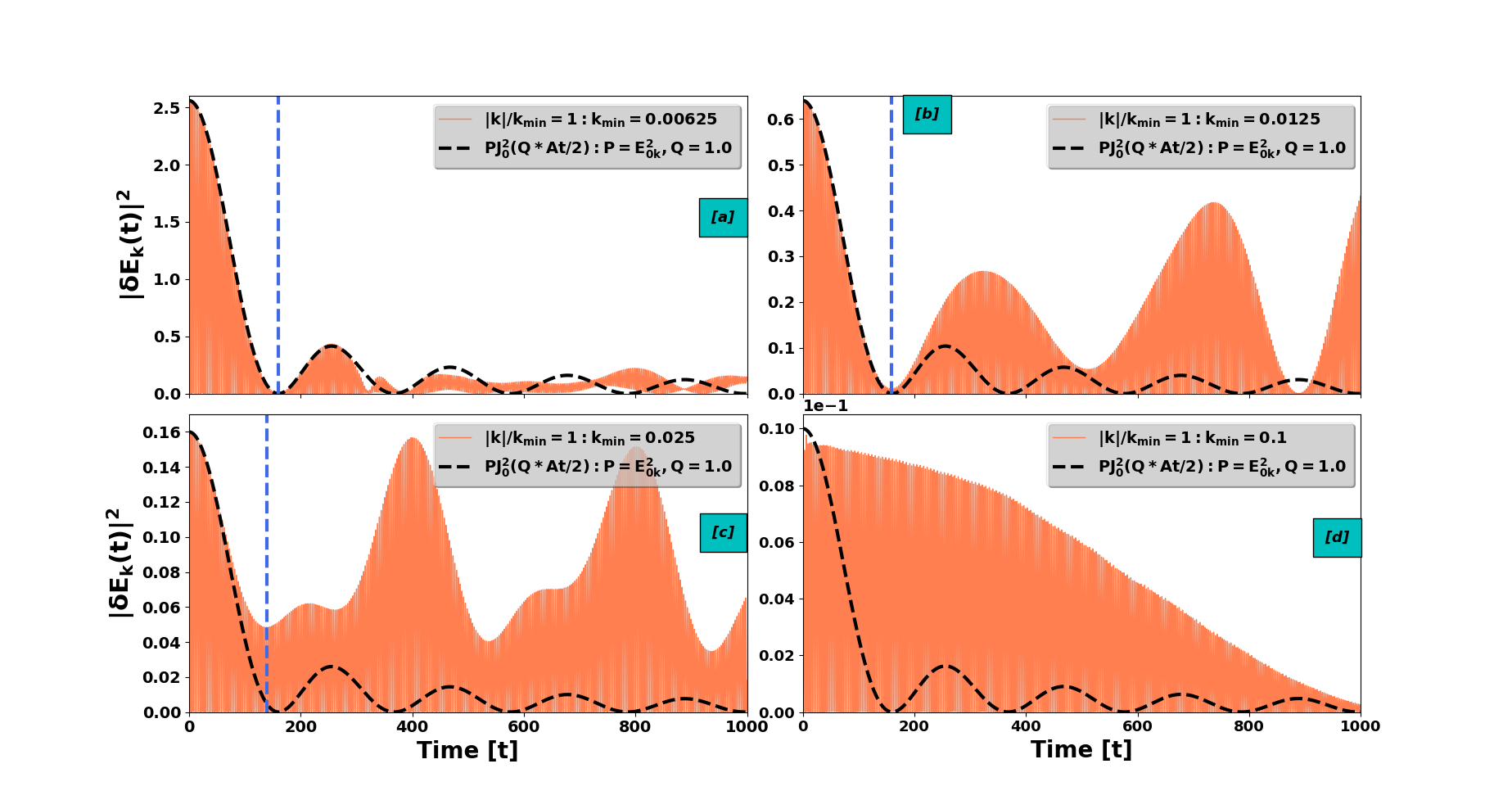}}
\caption{Temporal evolution of perturbation Fourier amplitudes $|\delta E_{k}|^{2}$ versus time $(t)$ obtained using kinetic (VPPM-OMP 1.0) solver for $(k_{0},k):(4,1),~\alpha=0.01,~A=0.03$ with various $k_{min}$ values i.e (a) $k_{min}=0.00625$, (b) $k_{min}=0.0125$, (c) $k_{min}=0.025$ and (d) $k_{min}=0.1$. Also, zeroth order Bessel function $PJ_{0}^{2}(QAt/2)$ with $P=E_{0k}^{2}:E_{0k}=\alpha/k$ and $Q=1.0$ is also plotted alongside each cases. It is evident from (a), (b), (c) and (d) that with the increase in the value of $k_{min}$ from 0.00625 to 0.1, Bessel $J_{0}^{2}(x)$ scaling strongly deviates from the perturbed Fourier amplitude $|\delta E_{k}|^{2}$.}
\label{41_K0K_EK2+J02_A=0.03_KMIN=0.00625+0.0125+0.025+0.1}
\end{figure*}

\begin{figure*}
\centerline{\includegraphics[scale=0.37]{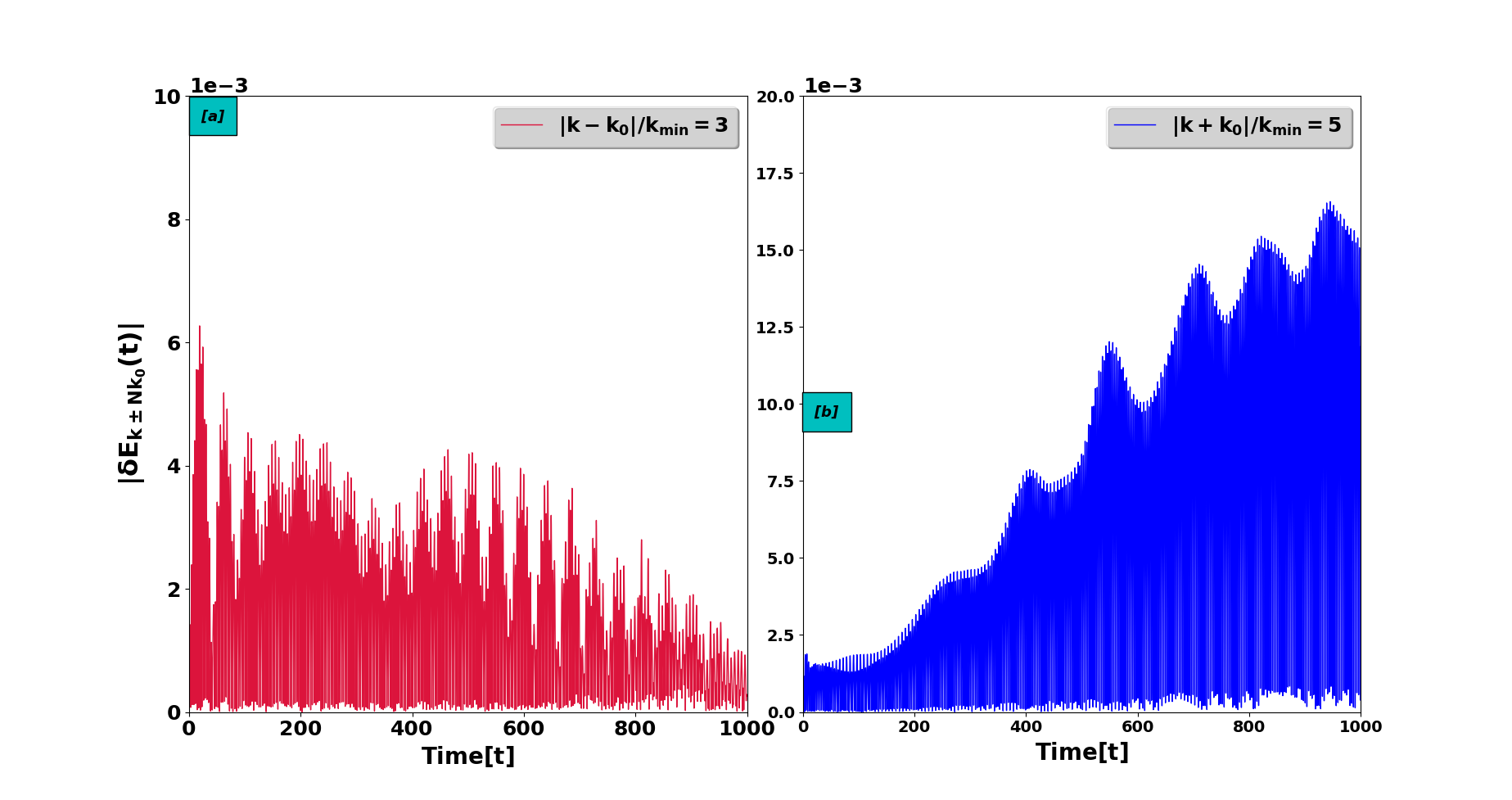}}
\caption{Temporal evolution of sideband Fourier amplitudes $|\delta E_{k \pm Nk_{0}}|:N \longrightarrow 1$ versus time $(t)$ obtained using kinetic (VPPM-OMP 1.0) solver for $(k_{0},k):(4,1),~\alpha=0.01,~A=0.03$ with $k_{min}=0.1$. Fig. \ref{41_K0K_EK2+J02_A=0.03_KMIN=0.00625+0.0125+0.025+0.1} (d) shows the primary perturbed mode $|\delta E_{k}|^{2}$ signature for this case.}
\label{41_K0K_EK_A=0.03_KMIN=0.1}
\end{figure*}

 In this Section, we will demonstrate the effect of decrease in the system size i.e $L_{max}=2\pi/k_{min}$ on the temporal evolution of the long-wavelength modes. In order to do so, with identical set of parameters as before i.e $(k_{0},k):(4,1),~\alpha=0.01,~A=0.03$, we increase the $k_{min}$ values from 0.00625 to 0.1. Fig. \ref{FLUID_41_K0K_EK2+J02_A=0.03_KMIN=0.00625+0.0125+0.025+0.1} and Fig. \ref{41_K0K_EK2+J02_A=0.03_KMIN=0.00625+0.0125+0.025+0.1} shows the variation of perturbation Fourier amplitudes $|\delta E_{k}|^{2}$ versus time $(t)$ obtained using fluid and kinetic solver for $(k_{0},k):(4,1),~\alpha=0.01,~A=0.03$ with various $k_{min}$ values i.e (a) $k_{min}=0.00625$, (b) $k_{min}=0.0125$, (c) $k_{min}=0.025$ and (d) $k_{min}=0.1$ alongwith zeroth order Bessel function $PJ_{0}^{2}(QAt/2)$ with $P=E_{0k}^{2}:E_{0k}=\alpha/k$ and $Q=1.0$ is also plotted alongside each cases. It is obivous from Fig. \ref{FLUID_41_K0K_EK2+J02_A=0.03_KMIN=0.00625+0.0125+0.025+0.1} [(a), (b), (c) and (d)] and Fig. \ref{41_K0K_EK2+J02_A=0.03_KMIN=0.00625+0.0125+0.025+0.1} [(a), (b), (c) and (d)] that with the increase in the $k_{min}$ values from 0.00625 to 0.1 i.e for more realistic system sizes $L_{max}$, the Bessel $J_{0}^{2}$ scaling does not hold true and strongly deviates from the perturbed electric field amplitude $|\delta E_{k}|^{2}$ solutions obtained either via fluid or via kinetic solvers. As reported by Kaw et al. \cite{kaw1973}, in the cold plasma limit i.e for $v_{\phi} \gg v_{thermal}$ with the progress of time, the energy density in amplitude of the primary perturbation mode $|\delta E_{k}|^{2}$ dies out as $J_{0}^{2}(At/2)$. It is obvious from Fig. \ref{FLUID_41_K0K_EK2+J02_A=0.03_KMIN=0.00625+0.0125+0.025+0.1} [(a), (b), (c) and (d)] that it is not true for all $k_{min}$ values, eventhough we are exactly solving the warm plasma fluid equations (as described in Sec. \ref{Fluid Model}) using our high resolution fluid (BOUT++) solver. Also, the deviations of $|\delta E_{k}|^{2}$ fluid solutions becomes more prominent with the increase in $k_{min}$ values. However, apart from qualitative similarity between fluid and kinetic numerical solutions for perturbed $|\delta E_{k}|^{2}$ mode with $k_{min}=0.00625,~0.0125,~0.025$ values [See Fig.  \ref{FLUID_41_K0K_EK2+J02_A=0.03_KMIN=0.00625+0.0125+0.025+0.1} (a), (b), (c) and Fig. \ref{41_K0K_EK2+J02_A=0.03_KMIN=0.00625+0.0125+0.025+0.1} (a), (b), (c)], there is a striking difference between the two for $k_{min}=0.1$ case [See Fig.  \ref{FLUID_41_K0K_EK2+J02_A=0.03_KMIN=0.00625+0.0125+0.025+0.1} (d) and Fig. \ref{41_K0K_EK2+J02_A=0.03_KMIN=0.00625+0.0125+0.025+0.1} (d)]. Also, the fluid and kinetic solutions of $k_{min}=0.025$ case is very interesting because the perturbed amplitude $|\delta E_{k}|^{2}$ regains all of its initial energy which is $E_{0k}^{2}=(\alpha/k)^{2}=(0.01/0.025)^{2}=0.16$ in subsequent cycles around $t=400~\omega_{pe}^{-1}$ as the coupling parameter being $N=1.0$ [See Fig.  \ref{FLUID_41_K0K_EK2+J02_A=0.03_KMIN=0.00625+0.0125+0.025+0.1} (c) and Fig. \ref{41_K0K_EK2+J02_A=0.03_KMIN=0.00625+0.0125+0.025+0.1} (c)].
 
   In Table \ref{TABLE 4}, we have tabulated frequency $(\omega)$, phase velocity $(v_{\phi}=\omega/k)$ and thermal velocity $(v_{thermal})$ for each perturbed mode $|\delta E_{k}|$ as shown in Fig. \ref{41_K0K_EK2+J02_A=0.03_KMIN=0.00625+0.0125+0.025+0.1} with various system scale i.e $k_{min}$ values. Even though for $k_{min}=0.1$ case the resonance conditions are not satisfied as $v_{\phi}>>v_{thermal}$, we see a damping of the perturbed electric field mode $|\delta E_{k}|$ at late times which is in contrast to the solutions obtained from fluid solver [See Fig.  \ref{FLUID_41_K0K_EK2+J02_A=0.03_KMIN=0.00625+0.0125+0.025+0.1} (d) and Fig. \ref{41_K0K_EK2+J02_A=0.03_KMIN=0.00625+0.0125+0.025+0.1} (d)]. Interestingly, investigation of the energy density in other coupled sideband modes $|\delta E_{k \pm Nk_{0}}|:N \longrightarrow 1$ as shown in Fig. \ref{41_K0K_EK_A=0.03_KMIN=0.1} [(a) and (b)], for $k_{min}=0.1,~\alpha=0.01,~A=0.03$ does not account for the damping in the total energy density of the primary perturbed mode $|\delta E_{k}|$. However, for intermediate $k_{min}$ values between 0.025 and 0.1, when we make transition from higher value (0.1) to lower value (0.025), the kinetic solutions qualitatively asymptotes to the fluid solutions.   

\begin{table}
\caption{ Oscillation frequency $[\omega]$, Phase velocity $[v_{\phi}=\omega/k]$ and Thermal velocity $[v_{thermal}]$ table for each perturbed mode $[k]$ with various system scale i.e $k_{min}$ values. For all $k_{min}$ values $v_{\phi} >> v_{thermal}$ i.e, the plasma is cold and hence no kinetic damping is expected.}  
\centering                         
\begin{tabular}{c c c c c}           
\hline\hline                        

System scale & Perturbation mode & Frequency & Phase Velocity & Thermal Velocity   \\    
$[k_{min}]$ & $[k]$ & $[\omega]$ & $[v_{\phi}]$ & $[v_{thermal}]$   \\ [1.0ex]    
\hline            
0.00625 & 0.00625 & 0.9801 & 156.816 & 1.0 \\          
0.0125 & 0.0125 & 0.9927 & 79.416 & 1.0 \\
0.025 & 0.025 & 0.9927 & 39.708 & 1.0 \\
0.1 & 0.1 & 1.0178 & 10.178 & 1.0  \\ [1ex]
\hline                               
\end{tabular}
\label{TABLE 4}
\end{table}

\begin{figure*}
\centerline{\includegraphics[scale=0.38]{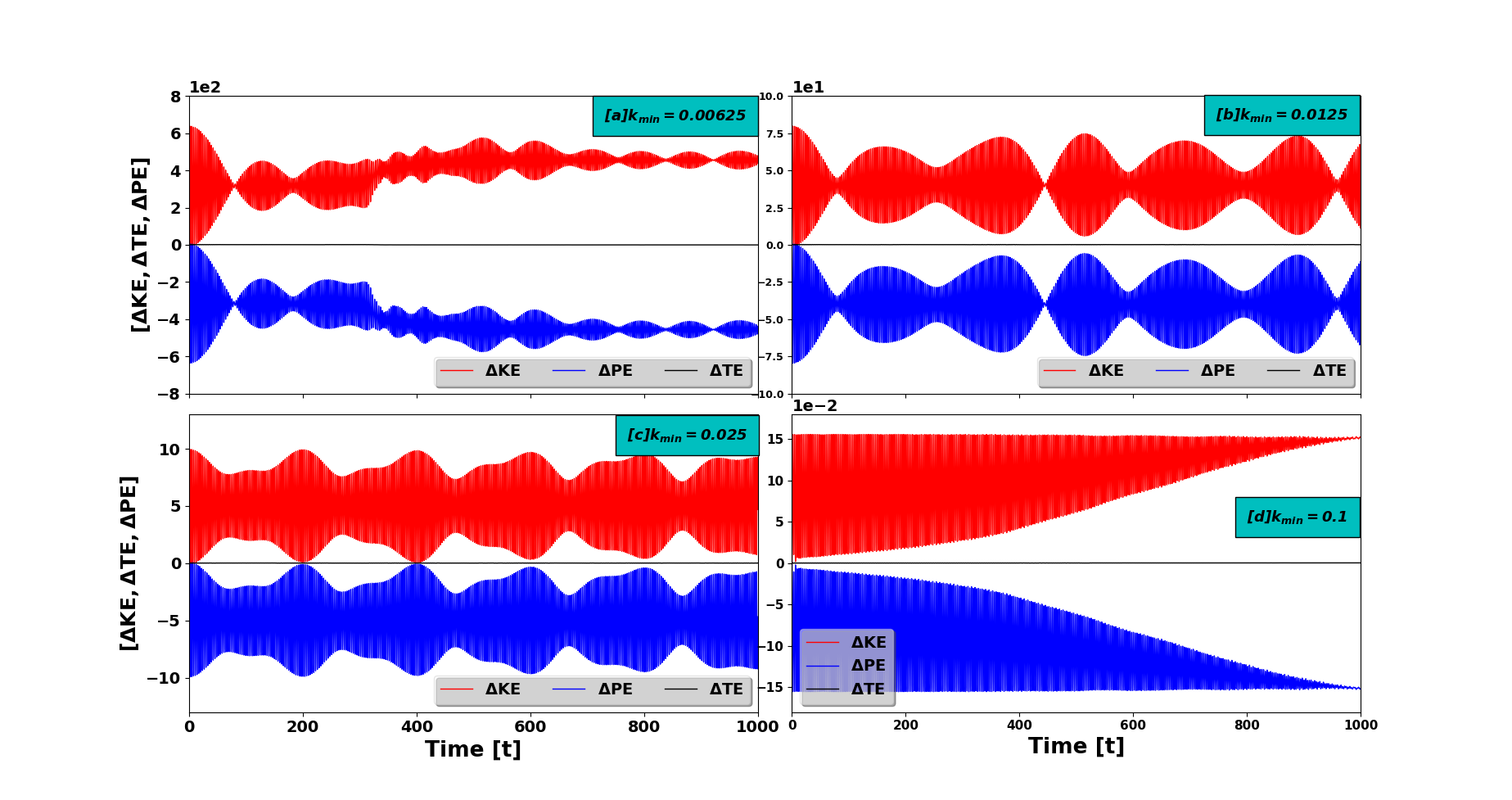}}
\caption{Relative kinetic, potential and total energies $(\Delta KE,~\Delta PE,~\Delta TE)$ variations with respect to time for solutions obtained using kinetic (VPPM-OMP 1.0) solver with $(k_{0},k):(4,1):k_{min}=0.00625,~\alpha=0.01,~A=0.03$ parameters alongwith various $k_{min}$ values i.e (a) $k_{min}=0.00625$, (b) $k_{min}=0.0125$, (c) $k_{min}=0.025$ and (d) $k_{min}=0.1$.  }
\label{41_K0K_EC_A=0.03_KMIN=0.00625+0.0125+0.025+0.1}
\end{figure*}

  To demonstrate the numerical consistency of the kinetic solutions obtained in Fig. \ref{41_K0K_EK2+J02_A=0.03_KMIN=0.00625+0.0125+0.025+0.1}, we have plotted the relative difference of kinetic energy $\Delta KE(t)=KE(t)-KE(0)$, potential energy $\Delta PE(t)=PE(t)-PE(0)$ and total energy $\Delta TE(t)=TE(t)-TE(0)$ defined by Eqs. \ref{EQ_KE}, \ref{EQ_PE} with respect to time for $(k_{0},k):(4,1):k_{min}=0.00625,~\alpha=0.01,~A=0.03$ parameters with various $k_{min}$ values i.e (a) $k_{min}=0.00625$, (b) $k_{min}=0.0125$, (c) $k_{min}=0.025$ and (d) $k_{min}=0.1$ as shown in Fig. \ref{41_K0K_EC_A=0.03_KMIN=0.00625+0.0125+0.025+0.1}.  
  
\begin{equation}
KE(t)=\int \int \frac{1}{2}v^{2}f(x,v,t)dxdv
\label{EQ_KE}
\end{equation}

\begin{equation}
PE(t)=\int \frac{1}{2}E^{2}(x,t)dx
\label{EQ_PE}
\end{equation} 

\begin{figure}
$~~~~~~~~~$
\includegraphics[scale=0.30]{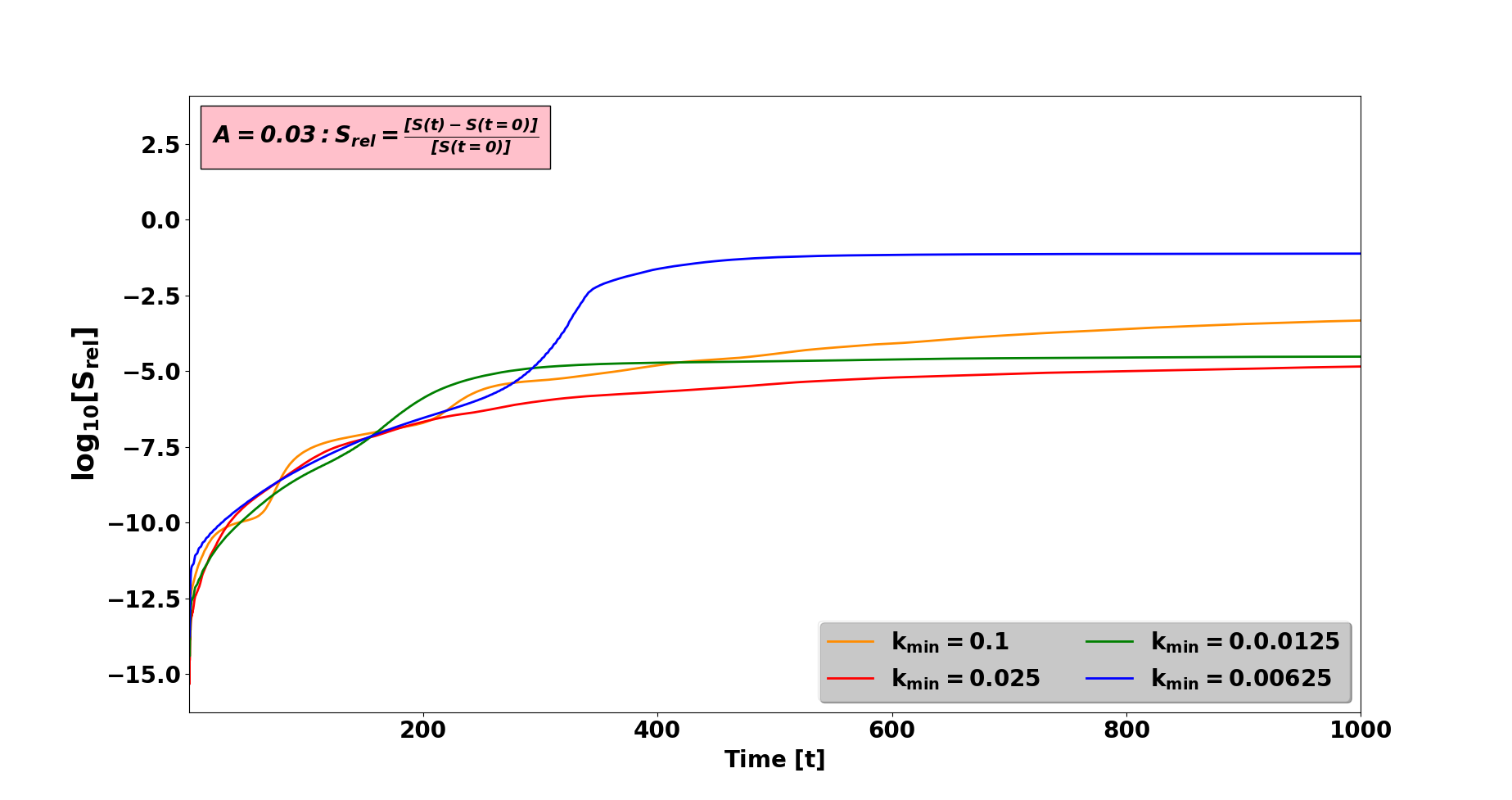}
\caption{Variation of relative entropy $S_{rel}$ given by Eq. \ref{EQ_SREL} with respect to time for solutions obtained using kinetic (VPPM-OMP 1.0) solver with $(k_{0},k):(4,1):k_{min}=0.006225,~\alpha=0.01,~A=0.03$ parameters alongwith various $k_{min}$ values i.e (a) $k_{min}=0.00625$, (b) $k_{min}=0.0125$, (c) $k_{min}=0.025$ and (d) $k_{min}=0.1$.}
\label{41_K0K_ENTROPY_A=0.03_KMIN=0.00625+0.0125+0.025+0.1 }
\end{figure}

    It indicates very good conservation and saturation of relative change in the energies with chosen grid discretization for each cases as tabulated in Table \ref{TABLE 1} i.e (a) $N_{x}=32768$, (b) $N_{x}=16384$, (c) $N_{x}=8192$ and (d) $N_{x}=2048$ respectively with $N_{v}=10000$ for both ions and electrons in space and velocity $(x,v)$ domains. Also, to demonstrates that the kinetic solutions we have obtained are steady state solutions, we show the signature of relative numerical entropy in Fig. \ref{41_K0K_ENTROPY_A=0.03_KMIN=0.00625+0.0125+0.025+0.1 } which is defined by Eqs. \ref{EQ_SREL} and \ref{EQ_ENTROPY} given as,

\begin{equation}
S_{rel}(t)=\frac{S(t)-S(t=0)}{S(t=0)}
\label{EQ_SREL}
\end{equation} 

\begin{equation}
S(t)=~-\int_{0}^{L_{max}} \int_{-v_{e}^{max}}^{+v_{e}^{max}} f(x,v,t) \log f(x,v,t)dxdv
\label{EQ_ENTROPY}
\end{equation} 
where $S(t)$ is the numerical entropy of the system. It acts as a measure of ``information lost" from the system as entropy $S(t)$ has tendency for monotonic increase with time due to the intrinsic ``filamentation'' property of Vlasov-Poisson system [\cite{manfredi1997,feix,sanjeev2021}]. Due to filamentation effect, the distribution function generates small scale structures in phase space $(x,v)$ which are dissipated when filamentation reaches the phase space grid size $[N_{x} \times N_{v} ]$ resolutions, leading to saturation in numerical entropy with time as shown in Fig \ref{41_K0K_ENTROPY_A=0.03_KMIN=0.00625+0.0125+0.025+0.1 }. It also signifies that information lost is small with respect to time for chosen grid size resolutions for each $k_{min}$ cases as tabulated in Table \ref{TABLE 1}, indicating a numerically stable, high quality simulation. 
\subsection{Fluid mode coupling with strong kinetic damping }
\label{Sepcial Case}

    As previously mentioned in Sec. \ref{Finite Size effects}, and shown in Fig. \ref{41_K0K_EK2+J02_A=0.03_KMIN=0.00625+0.0125+0.025+0.1}, inhomogeneous case with $k_{min}=0.1$ is a special case because we clearly observe the damping of the energy density of the primary mode $|\delta E_{k}|^{2}$ solutions obtained from kinetic solver, even though the resonance condition is not satisfied i.e $v_{\phi}>>v_{thermal}$ [Table. \ref{TABLE 4}] and the energy density in the coupled modes does not account for the loss in the energy of the primary mode [Fig. \ref{41_K0K_EK_A=0.03_KMIN=0.1}]. We have extended the simulations upto $t=3000~\omega_{pe}^{-1}$ with $(k_{0},k):(4,1),~k_{min}=0.1,~\alpha=0.01,~A=0.03$ parameter set, to investigate the involved kinetic effects associated with the coupled sideband modes which obviously can not be addressed via fluid solver.

\begin{figure}
$~~~~~~~~~$
\includegraphics[scale=0.30]{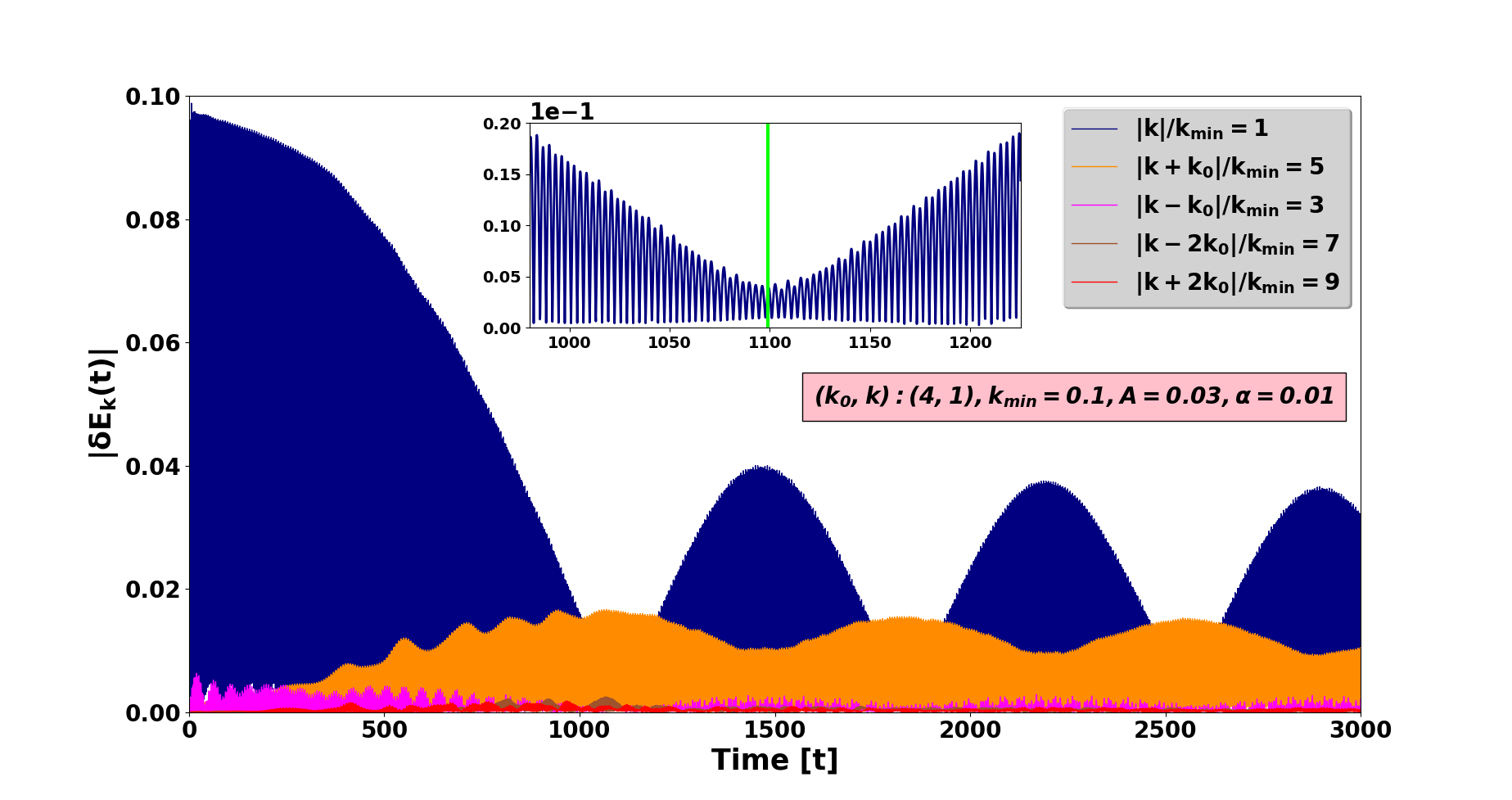}
\caption{Extended time evolution of perturbed primary and coupled secondary sideband electric field $|\delta E_{k}|$ modes obtained from kinetic solver VPPM-OMP 1.0 upto $t=3000~\omega_{pe}^{-1}$ with $(k_{0},k):(4,1),~A=0.03,~\alpha=0.01,~k_{min}=0.1,~N_{x} \times N_{v} = [2048 \times 10000]$. It illustrates the field signature of primary perturbation mode beyond $t=1000~\omega_{pe}^{-1}$ alongwith the late time growth in the coupled sideband modes i.e $k/k_{min}=3,~5$.  }
\label{41_K0K_EK_A=0.03_KMIN=0.1_N=2}
\end{figure}

\begin{table}
\caption{ Maximum oscillation frequency $(\omega_{k,s}^{max})$ obtained by 1D FFT (Fourier Transform) analysis and phase velocity $(v_{\phi,\omega_{s}}^{k}=\omega_{k,s}^{max}/k)$:$s \simeq 1-2,k \simeq 0.1-0.9$ table corresponding to the primary $[m^{th}]$ and secondary interacting $([m \pm n]^{th})$ sideband modes for $(k_{0},k):(4,1),~\alpha=0.01,~A=0.03$ case.}  
\centering                         
\begin{tabular}{c c c}           
\hline \hline                        
Mode No.$[m \pm n]$ & $[\omega_{k,s}^{max}]$ & $[v_{\phi,\omega_{s}}^{k}=\omega_{k,s}^{max}/k]$  \\ [1.0ex]    
\hline          
1 [m] & 1.0136 & 10.1360 \\          
2 [m+1] & 1.0580 & 5.290   \\ 
3 [m+2] & 1.0136 & 3.3780 \\
3 [m+2] & 1.1603 & 3.8676 \\
5 [m+4] & 1.0221 & 2.0442 \\
7 [m+6] & 1.3718 & 1.9597  \\
9 [m+8] & 1.7802 & 1.9781  \\ [1ex]
\hline                              
\end{tabular}
\label{TABLE 5}
\end{table}

   Fig. \ref{41_K0K_EK_A=0.03_KMIN=0.1_N=2} illustrates time evolution of perturbed primary and coupled secondary sideband electric field $|\delta E_{k}|$ modes obtained from kinetic solver upto $t=3000~\omega_{pe}^{-1}$ with $(k_{0},k):(4,1),~A=0.03,~\alpha=0.01,~k_{min}=0.1$. For this case, coupling parameter given by Eq. \ref{EQ_N}, as $N \sim \sqrt{0.03/3(0.4)^{2}} \sim 0.25$. It is clearly evident that coupled sideband modes $k/k_{min}=3,5$ gains some of the energy from the primary perturbed mode due to mode coupling phenemenon with background ion inhomogeneity which results into the late time growth in these sidebands. Also, there is negligible amount of energy transfer in the higher sidebands with $N=2$ i.e $k/k_{min}=7,9$. In order to investigate the resonance locations i.e phase velocities $(v_{\phi}=\omega/k)$ of each individual sideband modes we have tabulated maximum oscillation frequency $(\omega_{k,s}^{max})$ obtained by 1D FFT (Fast Fourier Transform) analysis and phase velocity $(v_{\phi,\omega_{s}}^{k}=\omega_{k,s}^{max}/k)$:$s \simeq 1-2,k \simeq 0.1-0.9$ corresponding to the primary $[m^{th}]$ and secondary interacting $([m \pm n]^{th})$ sideband modes for $(k_{0},k):(4,1),~\alpha=0.01,~A=0.03$ case given in Table \ref{TABLE 5}. Since, for primary perturbed mode $k/k_{min}=1$, phase velocity is 10.1360 which is out of the velocity domain ($v_{\phi} > v_{max}:v_{max}= \pm 6.0$), there will not be any kind of resonance wave particle interactions involved. However, resonance locations fall into bulk of the plasma ($v_{\phi}< \pm v_{max}: v_{max}= \pm 6.0$) for rest of the sideband modes, so, one can observe the vortex structure formation, even for the linear perturbation amplitudes as reported by \cite{sanjeev2021}.      

\begin{figure*}
\centerline{\includegraphics[scale=0.38]{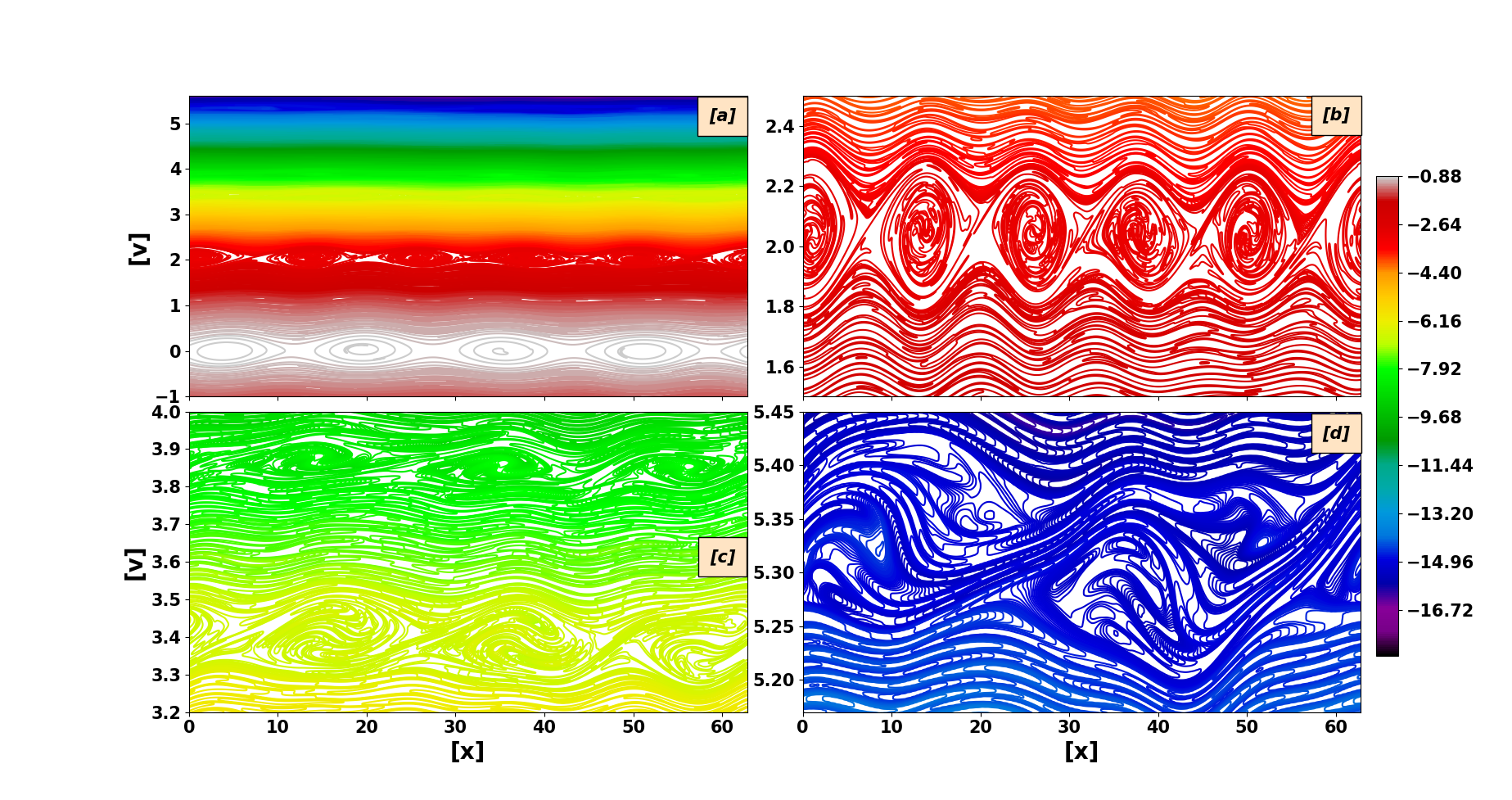}}
\caption{Phase space portrait of electron distribution function $f(x,v,t)$ at time $t=3000~\omega_{pe}^{-1}$ for $(k_{0},k):(4,1),~k_{min}=0.1,~A=0.03,~\alpha=0.01$ case. In (b), (c) and (d) zoomed plots are illustrated for various velocity ranges to show phase space vortices corresponding to $k/k_{min}=5,~3,~2$ modes formed at phase velocities tabulated in Table \ref{TABLE 5}.}
\label{41_K0K_CP_A=0.03_ALPHA=0.01_KMIN=0.1}
\end{figure*}

\begin{figure}
$~~~~~~~~~$
\includegraphics[scale=0.30]{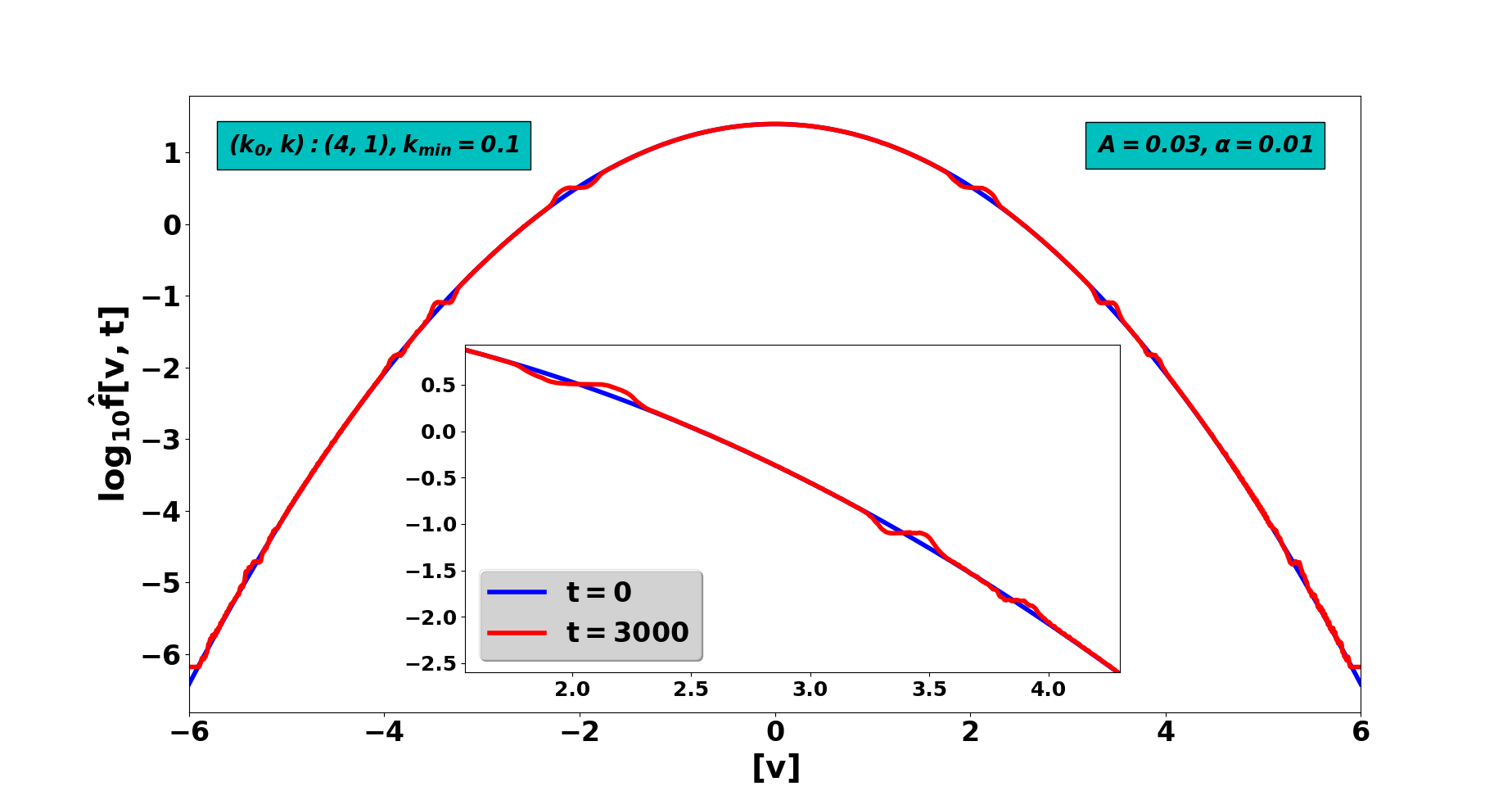}
\caption{Variation of the spatially averaged distribution function $\hat{f}(v,t)$ defined by Eq. \ref{EQ_DFE} with respect to velocity at $t=0,~3000~\omega_{pe}^{-1}$ for $(k_{0},k):(4,1),~k_{min}=0.1,~A=0.03,~\alpha=0.01$ case. It illustrates the plateau formation at velocity locations $v=2.0442,~3.3780,~3.8676,~5.2290$ which corresponds to the phase velocities of $k/k_{min}=5,~3,~2$ modes respectively [Table \ref{TABLE 5}].}
\label{41_K0K_DFE_A=0.03_ALPHA=0.01_KMIN=0.1}
\end{figure}
\begin{figure}
$~~~~~~~~~$
\includegraphics[scale=0.30]{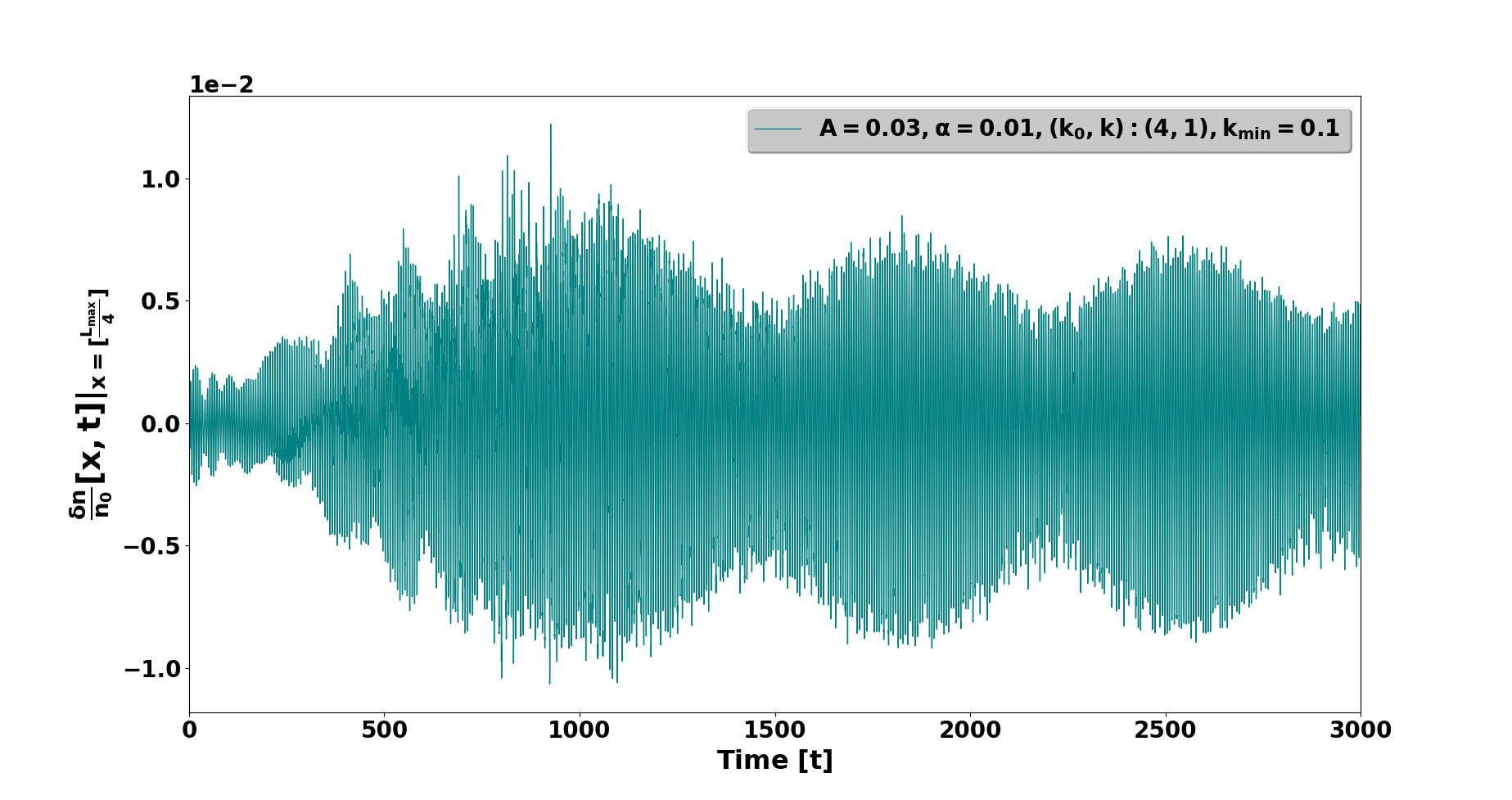}
\caption{Temporal evolution of excess density fraction at $x=L_{max}/4$ where $L_{max}=2\pi/k_{min}$ for $(k_{0},k):(4,1),~k_{min}=0.1,~A=0.03,~\alpha=0.01$ case. It demonstrates the particle trapping and detrapping modulations due to associated kinetic effects observed in coupled sideband modes.}
\label{41_K0K_EDF_A=0.03_KMIN=0.1}
\end{figure}

   Spatially averaged distribution function is given as,
   
\begin{equation}
\widehat{f}(v,t)= \frac{ \int_{0}^{L_{max}} f(x,v,t) dx }{ \int_{ +v_{e}^{max} }^{ -v_{e}^{max} } \int_{0}^{L_{max}} f(x,v,t) dx dv }
\label{EQ_DFE}
\end{equation}
where $f(x,v,t)$ is the evolved electron distribution function. Fig. \ref{41_K0K_CP_A=0.03_ALPHA=0.01_KMIN=0.1} nad Fig. \ref{41_K0K_DFE_A=0.03_ALPHA=0.01_KMIN=0.1} shows the portrait of electron distribution function $f(x,v,t)$ at time $t=3000~\omega_{pe}^{-1}$ and variation of the spatially averaged distribution function $\hat{f}(v,t)$ with respect to velocity at times $t=0,~3000~\omega_{pe}^{-1}$ respectively for $(k_{0},k):(4,1),~k_{min}=0.1,~A=0.03,~\alpha=0.01$ case. In Fig. \ref{41_K0K_CP_A=0.03_ALPHA=0.01_KMIN=0.1} [(b), (c) and (d)], due to resonance wave particle interaction phenomenon leading to particle trapping, one can observe the phase space vortex formation at $v=2.0442,~3.3780,~3.8676,~5.2290$ velocity locations which corresponds to the phase velocities of $k/k_{min}=5,~3,~2$ coupled sideband modes respectively as tabulated in Table \ref{TABLE 5}. Note that the phase space vortex chain at $v=0$ is due to equilibrium inhomogeneity $k_{0}/k_{min}=4$ with four periodic vortices [Fig. \ref{41_K0K_CP_A=0.03_ALPHA=0.01_KMIN=0.1} (a)]. Also, In Fig. \ref{41_K0K_DFE_A=0.03_ALPHA=0.01_KMIN=0.1}, the formation of the plateau or hump region in the spatially averaged distribution function around the phase velocity $v_{\phi}=\omega/k$ resonance locations indicate the particle trapping phenomenon due to involved kinetic effects \cite{manfredi1997}. Trapping of particles can be quantified by measuring an excess density fraction (EDF)$:(\delta n/n_{0})$ defined as a function of space and time in Eq. \ref{EQ_EDF} given as \cite{sanjeev2021,Pandey_2021_TPI_1,Pandey_2021_TPI_2},

\begin{equation}
\frac{\delta n}{ n_{0}}(x,t) ~ = ~ \left[ \frac{ n(x,t)-n(x,t=0)}{n(x,t=0)} \right]
\label{EQ_EDF}
\end{equation}

   Fig. \ref{41_K0K_EDF_A=0.03_KMIN=0.1} shows the temporal evolution of excess density fraction at $x=L_{max}/4$ where $L_{max}=2\pi/k_{min}$ for $(k_{0},k):(4,1),~k_{min}=0.1,~A=0.03,~\alpha=0.01$ case. It illustrates the particle trapping and detrapping modulations due to associated kinetic effects observed in coupled sideband $k/k_{min}=3,5$ modes. In the next section, we summarize and conclude the key results of the present work.


\section{Discussion and Conclusion}
\label{Discussion and conclusion}
  
  In this work, we have presented an extensive investigation of finite size and finite ion inhomogeneity amplitude effects on the dynamics of long-wavelength EPW modes in the presence of immobile non-uniform ion background using one-dimentional (1D), periodic kinetic (VPPM-OMP 1.0) and fluid (BOUT++) solvers. The important results are summarized as follows $: \longrightarrow$\\
  
\begin{itemize}
 \item Using our kinetic solver with high resolution simulations for inhomogeneous plasma $k_{min} = 0.00625,~(k_{0},k):(4,1),~A=0.03$ case, we have shown the damping of a long-wavelength EPW mode due to mode coupling phenomenon with inhomogeneous background of ions as predicted by Kaw et al. \cite{kaw1973}.
 
 \item Energy transfer process is shown between the coupled modes and its impact on the recovery of initial energy density of the primary perturbation mode i.e $|\delta E_{k}|^{2}$ in the subsequent cycles in a three and five mode coupled system.
 
 \item As stated by Kaw eta al. \cite{kaw1973} that in the cold plasma approximation i.e when $v_{\phi} \gg v_{the}$ with very large system size $L \rightarrow \infty$ or $k_{min} \rightarrow 0$, the energy density in the amplitude of EPW modes with wavenumber $|k \pm Nk_{0}|$ will evolve in time according to $N^{th}$ order Bessel function as $J^{2}_{N}(At/2)$.  Extensive comparative study is performed by solving both warm fluid and kinetic model equations, using fluid and kinetic solvers with identical parameter sets for various $(k_{min},~A)$ cases and we have demonstrated that the above prediction is only true for $k \longrightarrow 0$ i.e in the $v_{\phi}(=\omega/k) \gg v_{thermal}$ limit.
 
 \item Effect of ion inhomogeneity amplitude ($A$) on the temporal evolution of these long-wavelength EPW mode is investigated, which reveals that for a large ion amplitude inhomogeneity parameter sets, the zeroth order Bessel function $J_{0}^{2}(At/2)$ scaling estimates more cycles of amplitude evolution compared to the small ion inhomogeneity amplitudes in both solutions obtained either via kinetic or fluid solvers [Sec. \ref{Finite inhomogeneity amplitude effects}].
 
 \item Time required to attain first minimum of energy density signature $|\delta E_{k}|^{2}$ i.e $\tau_{FM}$ obtained for small value of $k_{min}$ i.e $k_{min}=0.00625$, scales with ion inhomogeneity amplitude ($A$) as $\tau_{FM} = [2Z_{J_{0}}/A:Z_{J_{0}}]$ where $Z_{J_{0}}=2.4048$ is the first zero of $J_{0}(x)$. The scaling and the $\tau_{FM}$ values obtained using both the kinetic and fluid solvers are exactly equal only for $k_{min}=0.00625$ case and it also matches the theoretical estimate of the energy transfer time $\tau_{ET}$ (sometimes also called phase mixing time) given by Kaw et al. \cite{kaw1973}.
 
 \item In contrast to the earlier studies, with the increase in inhomogeneity scale $k_{0}$, we have observed a significant shift in the $\tau_{FM}$ value compared to $\tau_{FM}^{k_{0}=0.025}$ and hence, we can conclude that $\tau_{FM}$ is dependent on both the inhomogeneity amplitude $A$ as well as inhomogeneity scale $k_{0}$ which is a interesting finding of this work.
 
 \item For finite realistic system sizes, with the increase in the $k_{min}$ values from 0.00625 to 0.1, the zeroth order Bessel function $J_{0}^{2}(At/2)$ scaling does not hold true in general and strongly deviates from the time evolution of energy density signature i.e $|\delta E_{k}|^{2}$ obtained either via kinetic or fluid solvers [Sec. \ref{Finite Size effects}]. 
 
 \item Despite choosing the exact parameter sets for comparative studies, following are the interesting differences between solutions obtained from kinetic (VPPM-OMP 1.0) and fluid (BOUT++) solvers $:\longrightarrow$\\
  \begin{itemize}
    \item For $k_{min}=0.00625,~A=0.03$ case, in kinetic solutions Bessel function $J_{0}^{2}(At/2)$ correctly maps the initial fall in energy density of the primary mode $|\delta E_{k}|^{2}$ upto $t=300~\omega_{pe}^{-1}$ whereas for fluid solutions the mapping is upto $t=360~\omega_{pe}^{-1}$. 
    
    \item Due to the absence of the involved kinetic effects, [Secs. \ref{Simulation Results Inhomogeneous plasma case with kmin}, \ref{Finite inhomogeneity amplitude effects}], late time amplitude surge from fluid model show in $|E_{k \pm Nk_{0}}|$ modes in solutions obtained from fluid solvers when compared to the kinetic ones.
    
    \item Although, the resonance conditions in $k_{min}=0.1$ case is not satisfied as $v_{\phi} \gg v_{thermal}$ (See. Table. \ref{TABLE 4}), we observe a damping in the perturbed energy density solution $|\delta E_{k}|^{2}$ obtained from kinetic solver at late times which is in contrast to the solutions obtained from fluid solver indicating the significance of the associated kinetic effects [Sec. \ref{Finite Size effects} and \ref{Sepcial Case}].
    
    \item In Sec. \ref{Sepcial Case}, to explain damping of the primary perturbed mode $|\delta E_{k}|^{2}$ signature for $k_{min}=0.1$ case, we have shown the phase space contour plots of electron distribution function $f(x,v)$ at $t=3000~\omega_{pe}^{-1}$, excess density fraction and spatially averaged distribution function $\hat{f}(v)$ for solutions obtained from kinetic solver with $A=0.03,~\alpha=0.01$, indicating the kinetic effects such as particle trapping, plateau formation etc, associated with sideband modes which can not be addressed via fluid solver.
        
    \item In obtaining $\tau_{FM}$ scaling with respect to $A$, with the gradual increase in the $k_{min}$ values from 0.00625 to 0.1, we observe that $\tau_{FM}$ obtained using fluid solver strongly deviates from the kinetic and scaling given by Eq. \ref{EQ_TAU_VS_A} i.e $[2Z_{J_{0}}/A]$ where $Z_{J_{0}}=2.4048$ is the first zero of $J_{0}(x)$ and a striking difference can be seen for $k_{min}=0.1$ case [Figs. \ref{41_K0K_TAU_ET_VS_A_KMIN=0_00625} and \ref{41_K0K_TAU_ET_VS_A_FOR_ALL_KMIN}].\\
    
  \end{itemize}
 
\end{itemize}

    Our study brings out several key aspects regarding temporal evolution of long-wavelength EPW modes which may be crucial in understanding the phenomena such as phase mixing, mode coupling and collisionless turbulence etc, in the presence of inhomogeneous equilibria which is relevant in laboratory as well as astrophysical plasmas. Investigations on temporal evolution of these EPW modes in the presence of mobile non-uniform ion background or in the presence of a spectrum of ion waves will be addressed in the near future.
    
\section*{Acknowledgments}
All numerical results of this paper were obtained using HPC cluster ANTYA at IPR. Authors would like to thank Jugal Chowdhury (IPR) for reading of the manuscript.

\section*{Data availability statement}
The data that support the findings of this study are available upon reasonable request from the authors.

\section*{References}
\bibliography{iopart-num}

\end{document}